\documentclass[acmtog]{acmart}
\usepackage{booktabs} 
\usepackage{bbm}
\usepackage{nicefrac}
\usepackage{algorithm}
\usepackage{algpseudocode}
\usepackage[normalem]{ulem}
\usepackage{empheq} 

\definecolor{codegreen}{rgb}{0,0.6,0}
\definecolor{codegray}{rgb}{0.5,0.5,0.5}
\definecolor{codepurple}{rgb}{0.58,0,0.82}
\definecolor{backcolour}{rgb}{0.96,0.96,0.98}
\usepackage{listings}
\usepackage{xcolor}
\setcitestyle{square}

\title{A Contact Proxy Splitting Method for Lagrangian Solid-Fluid Coupling} 

\author{Tianyi Xie}
\email{tianyixie77@g.ucla.edu}

\author{Minchen Li}
\email{minchernl@gmail.com}

\author{Yin Yang}
\email{yin.yang@utah.edu}

\author{Chenfanfu Jiang}
\email{chenfanfu.jiang@gmail.com}

\renewcommand{\shortauthors}{Xie et al.}

\usepackage{algorithm}
\usepackage{amsmath}

\DeclareMathOperator*{\argmin}{arg\,min}
\usepackage{algorithmicx}
\usepackage{algpseudocode}
\usepackage{multirow}
\usepackage{caption}
\usepackage{subcaption,bm}
\usepackage{ stmaryrd }
\algdef{SE}[DOWHILE]{Do}{doWhile}{\algorithmicdo}[1]{\algorithmicwhile\ #1}
\AtBeginDocument{%
  \providecommand\BibTeX{{%
    \normalfont B\kern-0.5em{\scshape i\kern-0.25em b}\kern-0.8em\TeX}}}

\newcommand{\mbf}[1]{\mathbf{#1}}
\newcommand{\csf}{C_\text{sf}}
\newcommand{\css}{C_\text{ss}}
\newcommand{\hcsf}{\hat{C}_\text{sf}}
\newcommand{\hcss}{\hat{C}_\text{ss}}

\begin{document}

\begin{abstract}
   We present a robust and efficient method for simulating Lagrangian solid-fluid coupling based on a new operator splitting strategy. We use variational formulations to approximate fluid properties and solid-fluid interactions, and introduce a unified two-way coupling formulation for SPH fluids and FEM solids using interior point barrier-based frictional contact. We split the resulting optimization problem into a fluid phase and a solid-coupling phase using a novel time-splitting approach with augmented \emph{contact proxies}, and propose efficient custom linear solvers. Our technique accounts for fluids interaction with nonlinear hyperelastic objects of different geometries and codimensions, while maintaining an algorithmically guaranteed non-penetrating criterion. Comprehensive benchmarks and experiments demonstrate the efficacy of our method.
\end{abstract}





\begin{teaserfigure}
  \includegraphics[width=\textwidth]{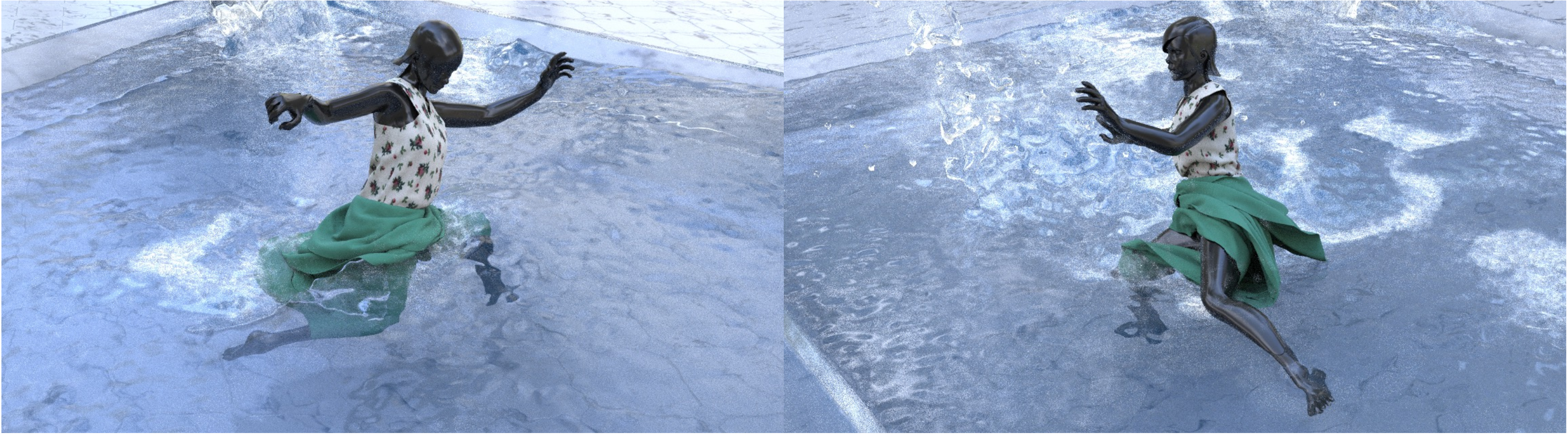}
  \caption{\textbf{Kick water.} Our method accurately captures the complex interactions between the water, the multi-layer skirt, and the mannequin body without any interpenetration as the mannequin wearing the skirt kicks in a swimming pool and sends water flying.}
  \label{fig: kick}
\end{teaserfigure}

\maketitle

\section{Introduction}
The coupling of solids and fluids is common  in nature but challenging to simulate. While solids are typically simulated using Lagrangian meshes, fluids are often discretized using Eulerian grids to accommodate topology changes. To accurately couple these distinct discretizations, sophisticated algorithms such as the method by \citet{zarifi2017positive} are often necessary. Unfortunately, they can be expensive and do not handle thin shells. 
Purely Eulerian \cite{teng2016eulerian, valkov2015eulerian} or SPH \cite{akinci2012versatile, gissler2019interlinked} schemes have demonstrated successful two-way coupling by same-view discretization. They do not easily extend to nonlinear elastodynamics. Hybrid methods like MPM \cite{jiang2016material} can simulate mixed materials, but can experience artificial stickiness unless resolved with more expensive schemes \cite{fang2020iq}. Furthermore, these methods do not ensure non-intersecting trajectories and often require additional correction procedures to handle accidentally penetrated fluids during advection.


 We take the Lagrangian path and present a new method for coupling FEM solids and SPH fluids. By approximating solid, fluid, and interaction terms with potentials, we formulate two-way coupling as an optimization problem.
Specifically, we draw inspiration from position-based fluids \cite{macklin2013position} and model weak incompressibility using a quadratic energy and a new \emph{updated Lagrangian update rule} to track volume changes. We further symmetrize the discrete Laplacian-based viscosity and propose a \emph{discrete quadratic potential} for better accuracy and robustness. We follow the Incremental Potential Contact \cite{li2020incremental} model to enforce guaranteed separable boundary conditions and resolve frictional contacts at the interface.

The proposed formulation achieves strong coupling, but can be exceedingly inefficient when solved with Newton's method due to the huge and dense Hessian of the fluid part, which is a result of the need for many particle neighbors for an accurate SPH discretization. This causes a significant computing bottleneck.

To tackle this issue, we propose a robust  \emph{proxy contact energy} formulation, spitting the time integration into a fluid phase and a solid-coupling phase. The fluid phase requires only one Newton iteration per time step, resulting in increased efficiency with nonlinear optimization occurring only during the solid-coupling phase. 
One of the key advantages of our quadratic  proxy is its ability to effectively resolve instability caused by time splitting. This is achieved through its asymptotic approximation to the solid-fluid contact force. Additionally, time integration is maintained consistent through the cancellation of the proxy's contribution in the solid-coupling phase, resulting in only a small splitting error.
Finally, we design a matrix-free conjugate gradient solver and a domain-decomposed solver to further enhance the computational efficiency.


\section{Related Work}

Traditional fluid solvers typically use an Eulerian grid. Many existing works focus on coupling Eulerian fluids with Lagrangian solids by resolving interactions between the grids and irregular mesh boundaries. The ghost fluid method \cite{fedkiw1999non,fedkiw2002coupling} was proposed to additionally discretize the Eulerian/Lagrangian interface. 
Early works considered weak coupling \cite{guendelman2005coupling}, which advances the solids and fluids alternatively. Strong coupling \cite{klingner2006fluid} on the other hand solves a monolithic system and is often more robust. 
The cut-cell method\ \cite{roble2005cartesian} is another widely used solution, often through the usage of virtual nodes.
\citet{batty2007fast} proposed a variational framework to strongly couple fluids and rigid bodies by casting the pressure solve into minimization. Subsequent extensions support deformable objects and thin shells\ \cite{robinson2008two,robinson2011symmetric}, where  elastic forces are explicitly applied and the coupling step is implicit. Assuming corotated linear elasticity, \citet{zarifi2017positive} incorporated the implicit solid dynamics into pressure projection, obtaining a symmetric positive-definite system. Later works also explored rigid-rigid \cite{takahashi2020monolith} and rigid-fluid \cite{takahashi2021frictionalmonolith} frictional contacts. 
Eulerian solids \cite{levin2011eulerian} were also explored, where coupling can be conveniently achieved in a purely Eulerian fashion \cite{teng2016eulerian, valkov2015eulerian}. However they face challenges in  numerical dissipation, volume conservation, and handling structures thinner than a grid cell. More recently, \citet{brandt2019reduced} built upon the immersed boundary method \cite{peskin2002immersed} and proposed a reduced solver to simulate real-time coupling, focusing on  incompressible elastic materials and no-slip boundary conditions.

Fluids can also be directly modeled with Lagrangian meshes \cite{clausen2013simulating, klingner2006fluid,wang2020codimensional,batty2012discrete}, enabling explicit coupling with solids. However, remeshing tends to become a  bottleneck. Using particles is another popular strategy. 
SPH \cite{koschier2022survey} uses spatial sampling to approximate continuous functions and has been shown compelling for fluid dynamics. Pioneering works \cite{monaghan1994simulating, becker2007weakly} used the Equation of State (EOS) for weakly compressible fluids, where the pressure is proportional to the density deviation. An explicit formulation may strictly restrict time step sizes. Incompressibility has also been enforced by solving a Pressure Poisson Equation (PPE) \cite{solenthaler2009predictive,ihmsen2013implicit,bender2015divergence}. This approach seeks to cancel out density or velocity divergence deviations caused by non-pressure forces through the use of pressure accelerations.
SPH boundary handling techniques have been developed to prevent penetrations of fluid particles near solid boundaries \cite{becker2009direct, ihmsen2010boundary,becker2007weakly}. One such method, proposed by \citet{akinci2012versatile}, uses a single layer of boundary samples and has been applied to the coupling of fluids with both rigid bodies and elastic solids \cite{akinci2013coupling}.
\citet{gissler2019interlinked} proposed a global formulation that unifies rigid body and fluid dynamics, in which the fluid pressure solver is linked to a second artificial pressure solver for rigid body particles. \citet{koschier2017density} introduced an alternative method using density maps to represent dynamic rigid boundaries, eliminating the need for boundary particles. \citet{bender2019volume} proposed using the volume contribution of boundary geometry to compute boundary forces, which reduces the cost of precomputation but cannot be applied to deformable bodies.

\citet{solenthaler2007unified} used SPH to approximate the deformation gradient of linear elastic materials, but the resulting gradient is not rotation invariant. \citet{becker2009corotated} addressed this issue by using shape matching to determine orientation and calculating forces in a rotated configuration. \citet{peer2018implicit} proposed an implicit scheme and applied kernel gradient correction \cite{bonet1999variational} to obtain a first-order consistent SPH formulation for the deformation gradient.
Incorporating solid particles into the preexisting fluid pressure solver can resolve contact handling, but SPH still faces numerical issues such as the zero-mode \cite{kugelstadt2021fast, ganzenmuller2015hourglass} when simulating elastic objects. Additionally, the pressure solver will treat solid objects as incompressible under compression, which may not be applicable in all cases.

The Material-Point Method (MPM) \cite{sulsky1995application, jiang2016material} combines Lagrangian and Eulerian representations to capture solid-fluid coupling \cite{stomakhin2014augmented,fei2018multi,yan2018mpm} and mixture \cite{tampubolon2017multi,gao2018animating}. Fang et al. [\citeyear{fang2020iq}] proposed a free-slip treatment, but did not consider separation. 
Recently, a FEM-MPM coupling method based on a variational barrier formulation \cite{li2020incremental} has been proposed for coupling frictional and separable elastic materials \cite{li2022bfemp}. Our approach for solid-fluid coupling is inspired by this method and uses a similar purely Lagrangian framework.

\begin{figure}[t]
  \centering
  \includegraphics[width=\linewidth]{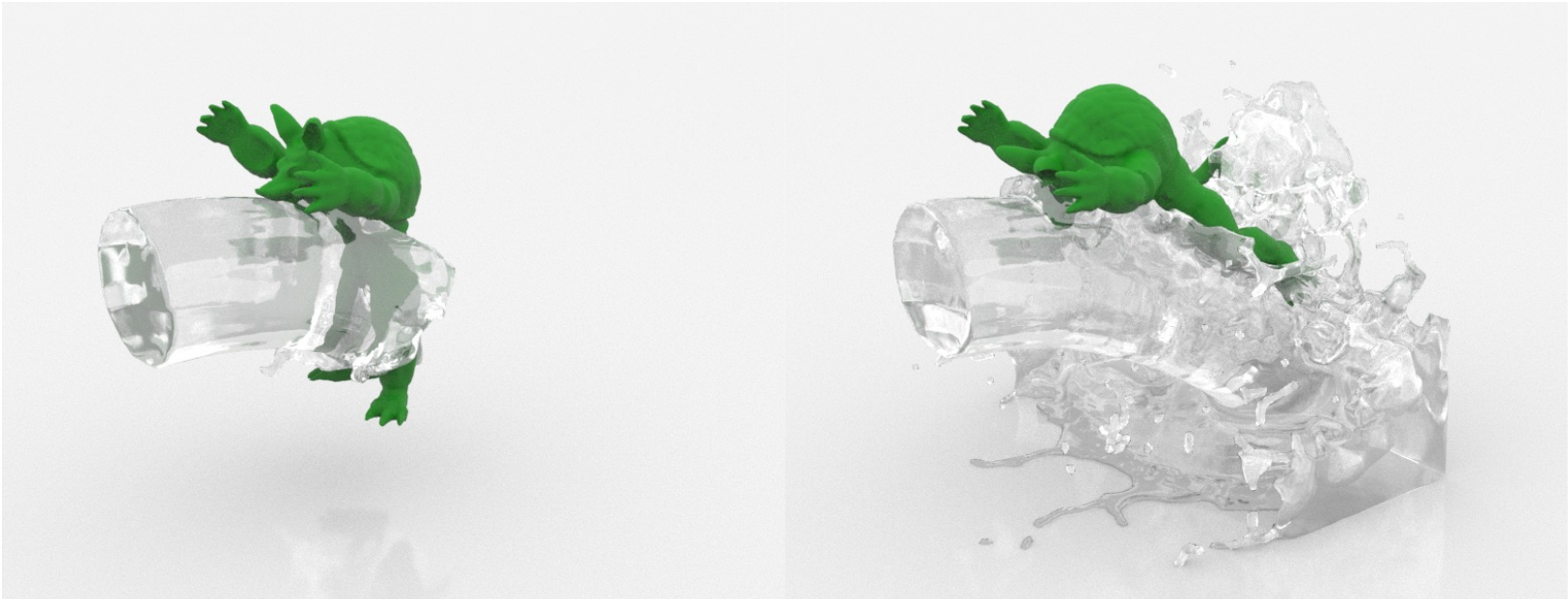}
  \caption{\textbf{Shooting armadillo} with a high-speed water jet.}
  \label{fig: shot armadillo}
\end{figure}

\section{Formulation}
Here we derive a time integrator for a coupled system of solids and fluids by starting with the governing equations and then performing discretization. Subscripts $s$ and $f$ represent solid and fluid quantities.

\subsection{Governing Equations}
The governing equations for the coupled system  are
\begin{align}
    \rho_s \frac{D \mbf{v}_s}{D t} &= \nabla\cdot \mbf{\mathbf{\sigma}} + \rho_s \mbf{g} + \mathbf{f}_{s\shortrightarrow s} + \mathbf{f}_{f\shortrightarrow s}, \label{eq: solid momentum} \\
      \rho_f \frac{D \mbf{v}_{f}}{D t} &= -\nabla p + \mu \nabla^2 \mbf{v}_f + \rho_f \mbf{g} -\mathbf{f}_{f\shortrightarrow s}, \label{eq: navier}\\
    \nabla \cdot \mbf{v}_f &= 0, \label{eq: div_free}
\end{align}
where $\rho$ is density, $\mbf{g}$ is gravity, $\mathbf{f}_{s\shortrightarrow s}$ is the self-contact force of solids, $\mathbf{f}_{f\shortrightarrow s}$ is the contact force exerted by fluids,  $\mathbf{\sigma}$ is Cauchy stress, $p$ is pressure and $\mu$ is the dynamic viscosity\ \cite{bridson2015fluid}.
  

At the interface between solids and fluids, we enforce the separable boundary condition
\begin{equation}
    0 \leq (\mbf{v}_s - \mbf{v}_f) \cdot \mbf{n}_f \ \bot \
    (\mathbf{f}_{f\shortrightarrow s} \cdot \mbf{n}_f) \geq 0
    \label{eq: boundary condition}
\end{equation}
to prevent penetration while allowing separation \cite{batty2007fast}. This condition helps determine the normal component of $\mathbf{f}_{f\shortrightarrow s}$. 
For the tangential component (friction), let $\mathbf{u} = (\mbf{I} - \mbf{n}_f \otimes \mbf{n}_f)(\mbf{v}_s - \mbf{v}_f)$ be the tangential relative velocity, we have
\begin{equation}
    \begin{gathered}
        (\mbf{I} - \mbf{n}_f \otimes \mbf{n}_f) \mathbf{f}_{f\shortrightarrow s} = \argmin_{\bm{\beta}}\bm{\beta} \cdot \mathbf{u} \\
        \text{s.t.} \quad \|\bm{\beta}\|\leq \mu_t \mathbf{f}_{f\shortrightarrow s} \cdot \mbf{n}_f \ \ \text{and} \ \ \bm{\beta} \cdot \mbf{n}_f = 0
    \end{gathered}
\end{equation}
following the Maximum Dissipation Principle\ \cite{moreau2011unilateral}, where $\mu_t$ is the friction coefficient. 
We enforce exact mass conservation by adopting Lagrangian methods to discretize both domains.

\subsection{Solid Domain}
We focus on nonlinear hyperelastic solids, where the elastic force is the negative gradient of an elastic potential.
After discretizing the solid domain $\Omega_s$ as Lagrangian linear finite elements (triangles in 2D and tetrahedra in 3D), the total elastic potential is a piecewise constant summation of an elastic energy density function $\psi_s(\mbf{F})$ (e.g. neo-Hookean) over the mesh domain: $    \Psi_s(\mbf{x}) = \sum_{e} V_e \psi_s(\mbf{F}_e)$,
where $V_e$ is the rest volume of tetrahedron $e$, and $\mbf{F} = \frac{\partial \mbf{x}(\mbf{X}, t)}{\partial \mbf{X}}$ is the deformation gradient with $\mbf{X}$ and $\mbf{x}$ the material and world space coordinates respectively \cite{sifakis2012fem}.
For $\mbf{f}_{s\shortrightarrow s}$, we follow \citet{li2020incremental}'s smooth barrier approach that guarantees non-penetration. We leave the discussion of $\mbf{f}_{f\shortrightarrow s}$ to \S~\ref{sec:coupling}.

\subsection{Fluid Domain}
Following SPH literature \cite{becker2007weakly, macklin2013position, ihmsen2013implicit, bender2015divergence}, we discretize the fluid domain $\Omega_f$ with Lagrangian particles. 
To integrate fluids with optimization-based time integration, we approximate both the pressure and viscosity forces as conservative forces. We verify in the appendix that these proposed potential energies are both convex and quadratic.

\subsubsection{Incompressibility Potential}
Pressure forces help preserve the volume of incompressible fluids. 
We thus model the incompressibility via a quadratic energy density function $\psi_{f, I}(J) = \frac{k_I}{2}(J-1)^2$ that penalizes the deviation of volume ratio $J = \rho_0/\rho$ from $1$, where $\rho_0$ is the initial density. The use of a large stiffness value ($k_I$) in a convergent solve results in negligible visual compression, eliminating the need for higher degree polynomials in nearly incompressible fluids \cite{hyde2020implicit}. The incompressibility potential is obtained by integrating $\psi_{f, I}(J)$ over the fluid domain $\Omega_f^0$ in \emph{material space}:
\begin{equation}
    P_I(\mbf{x}) = \sum_i \frac{k_I}{2} V_0(J_i(\mbf{x}) - 1)^2,
\end{equation}
where we assumed all fluid particles have equal rest volume $V_0$, and $J_i$ denotes the volume ratio of the $i$-th particle as a function of $\mbf{x}$.

\paragraph{Updated Lagrangian} SPH literature often relate $\rho$ to $\mbf{x}$ through density summation in the world space. 
To obtain a linear relation between $J$ and $\mbf{x}$ so that the incompressibility potential stays quadratic in terms of $\mbf{x}$, we track $J$ in an \emph{updated Lagrangian} fashion. Treating $\Omega^n$ as an intermediate reference space and differentiating the deformation map between $\Omega^n$ and $\Omega^{n+1}$ results in an update rule   
\begin{equation}
    J_i^{n+1} = J^n_i ( 1 + h \nabla \cdot \mathbf{v}^{n+1}_i),
\end{equation}
where $J_i^n$ and $\nabla \cdot \mathbf{v}^{n+1}_i$ can be approximated as
\begin{align}
    J_i^n =  \frac{\rho_0}{\sum\limits_{j} m_j W_{ij}}, \ 
    \nabla \cdot \mathbf{v}^{n+1}_i = \sum_j \frac{m_j}{\rho^{n}_j} (\mathbf{v}^{n+1}_j - \mathbf{v}^{n+1}_i) \cdot \nabla_i W_{ij} \label{eq: sph div vel approx}
\end{align}
via SPH, and $W_{ij} = W(\mbf{x}_i - \mbf{x}_j)$ is a kernel function (e.g. Cubic Spline kernel \cite{monaghan1992smoothed, monaghan2005smoothed} or Spiky kernel \cite{muller2003particle}). 
Here $J^n_i$ denote the reinitialized volume ratio of the $i$-th fluid particle at the beginning of time step $n$. Such reinitialization avoids accumulated density and particle distribution errors commonly seen in other updated Lagrangian solvers like MPM.
 


\subsubsection{Viscosity Potential}
\begin{figure}[t]
  \centering
  \includegraphics[width=\linewidth]{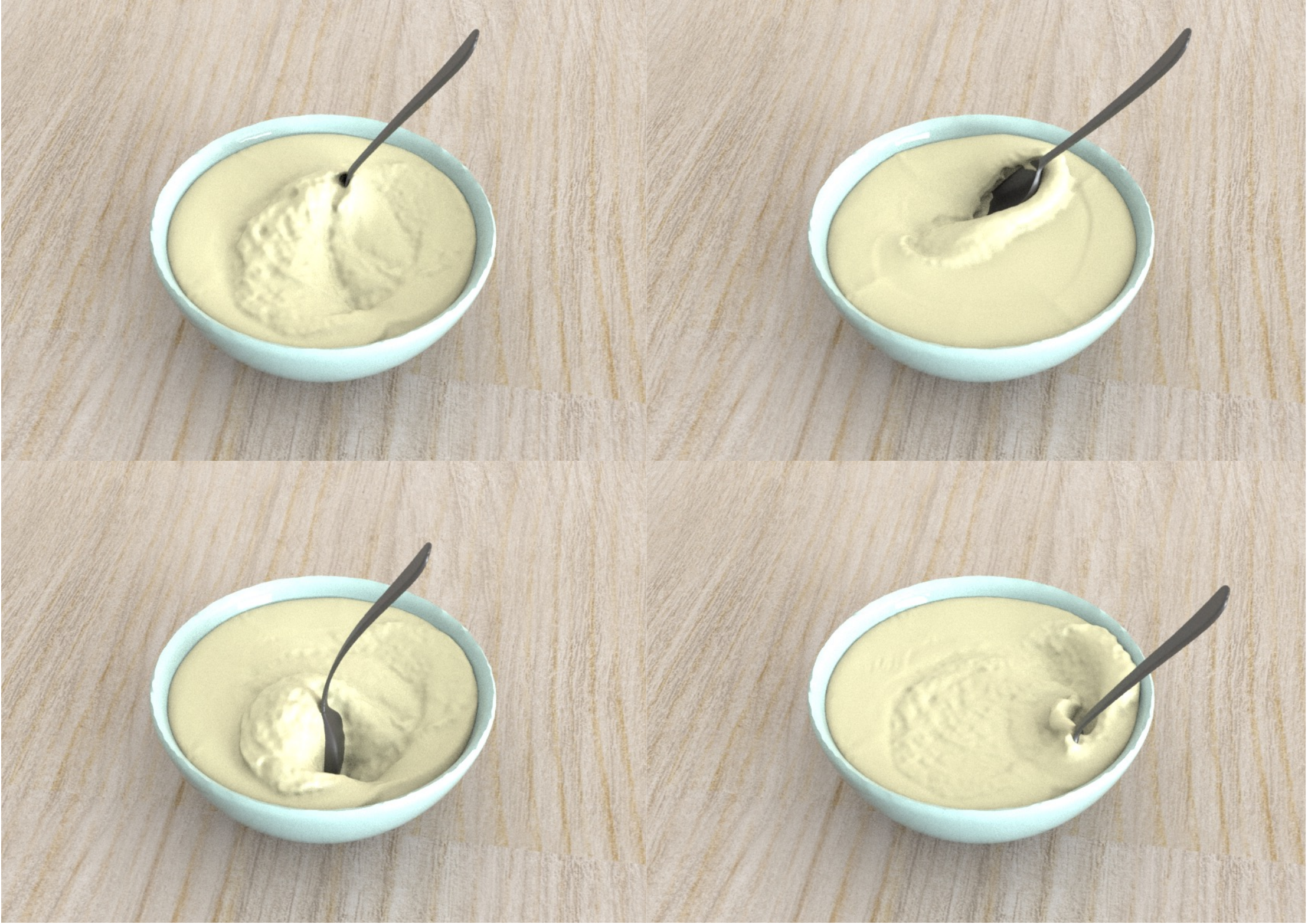}
  \caption{\textbf{Cream} is stirred, causing spoon deformation.}
  \label{fig: cream}
\end{figure}
Modeling viscosity via strain rate tensors \cite{peer2015implicit, peer2016prescribed, takahashi2015implicit, bender2016divergence} is possible, but may suffer from artifacts at the surface due to particle deficiencies. We follow \citet{monaghan2005smoothed} to use the more robust velocity Laplacian \cite{weiler2018physically} and derive its energy form.
\begin{figure*}[t]
  \centering
  \includegraphics[width=\linewidth]{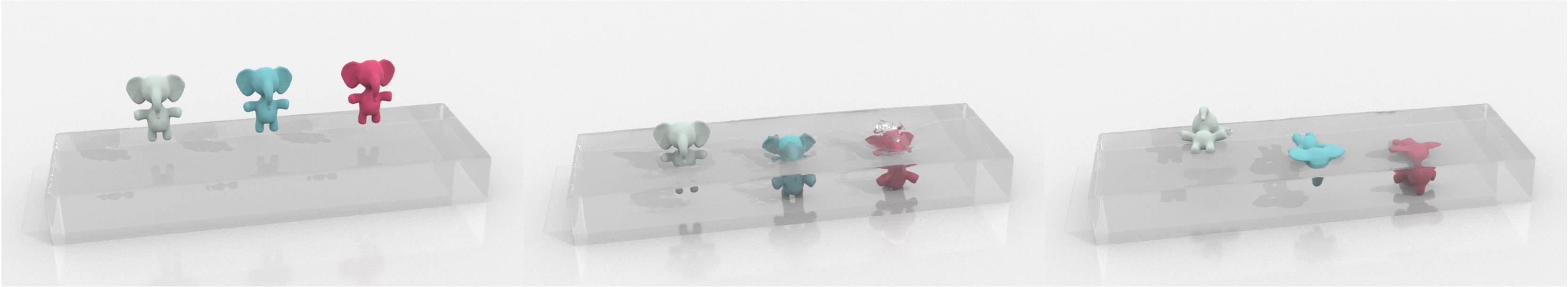}
  \caption{\textbf{Buoyancy}. Three elastic elephants with different densities (from left to right: 200, 700, and 1200 $kg/m^3$) fall into the water, demonstrating buoyancy.}
  \label{fig: buoyancy}
\end{figure*}
Combining SPH 1st-order derivatives and finite differences, the viscosity force can be computed as
\begin{align*}
    \mathbf{f}_i(\mathbf{x}) &= \nu m_i \nabla^2 \mathbf{v}_i^{n+1}
          = 2 \nu (d+2)\sum_j \frac{m_i m_j}{\rho_j}\frac{\nabla_i W_{ij} (\mathbf{x}^n_{ij})^T}{\|\mathbf{x}_{ij}^n\|^2 + 0.01 \hbar^2} \mathbf{v}_{ij}^{n+1},
\end{align*}
where $\mathbf{x}^n_{ij} = \mathbf{x}^n_i - \mathbf{x}^n_j$, $\mathbf{v}^{n+1}_{ij} = \mathbf{v}^{n+1}_i - \mathbf{v}^{n+1}_j$, $\hbar$ is the support radius of the kernel, $\nu$ and $d\in\{2,3\}$ denote the kinematic viscosity and spatial dimension respectively. Directly applying this force violates momentum conservation as the mutual interaction forces are not equal and opposite. Thus, we perform a further approximation
\begin{equation}
\begin{aligned}
    \mathbf{f}_i(\mathbf{x}) &\approx 4\nu (d+2)\sum_j \frac{m_i m_j}{\rho_i + \rho_j}\frac{\nabla_i W_{ij} (\mathbf{x}^n_{ij})^T}{\|\mathbf{x}_{ij}^n\|^2 + 0.01\hbar^2} \mathbf{v}_{ij}^{n+1}
\end{aligned}
\end{equation}
to solve this issue and also make the force integrable.
Let $ \mathbf{V}_{ij} = 4 (d+2) \frac{m_i m_j}{\rho_i + \rho_j}\frac{(-\nabla_i W_{ij} )(\mathbf{x}^n_{ij})^T}{\|\mathbf{x}_{ij}^n\|^2 + 0.01\hbar ^2}$, we can now gather and integrate all viscosity forces and obtain a quadratic viscosity potential
\begin{equation}
\begin{aligned}
    P_V(\mathbf{x}) &= \frac{1}{4} \nu \hat{h} \sum_i \sum_j \|\mathbf{v}^{n+1}_{ij}\|^2_{\mathbf{V}_{ij}},
\end{aligned}
\end{equation}
where $\hat{h}$ is a constant scalar related to the time integration scheme. For example, $\hat{h} = h$ for implicit Euler as $\mathbf{v}^{n+1} = (\mathbf{x}^{n+1}-\mathbf{x}^n)/h$.

\begin{figure*}[t]
  \centering
  \includegraphics[width=\linewidth]{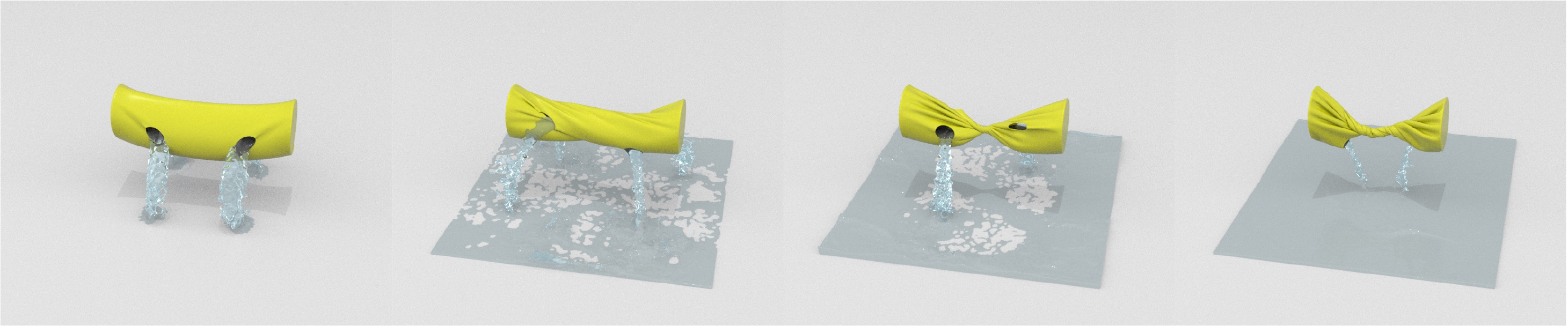}
  \caption{\textbf{Twist cylinder}. A cylindrical cloth with four holes is twisted, squeezing out water from the inside.}
  \label{fig: twist cylinder}
\end{figure*}

\subsection{Coupling} \label{sec:coupling}
\subsubsection{Barrier Potential for Non-penetration}
To couple the solid domain $\Omega_s$ with the fluid domain $\Omega_f$, we use the separable boundary condition (Eq. \ref{eq: boundary condition}), which enforces non-interpenetration constraints between these two domains. To model these constraints, we first define a distance function
\begin{equation}
    d(\partial\Omega_s^t, \mbf{x}_f) = \min_{\mbf{x}_s} \|\mbf{x}_s - \mbf{x}_f\|, \quad \mbf{x}_s \in \partial\Omega_s^t, \ \mbf{x}_f \in \Omega_f^t,
\end{equation}
which measures the distance between $\mbf{x}_f$, a point in the fluid domain, and the surface of the solid domain. Then the primal component of the constraints can be expressed as
\begin{equation}
    d(\partial\Omega_s^t, \mbf{x}_f) \geq 0, \quad \forall t \geq 0, \ \forall \mbf{x}_f \in \Omega^t_f.
\end{equation}
We then adopt the barrier formulation from \citet{li2020incremental} to model all the constraints in Eq. \ref{eq: boundary condition} between solids and fluids, and obtain a barrier potential
\begin{equation}
    \int_{\partial \Omega^0_f} b (d(\partial\Omega_s^t, \mbf{x}_f), \hat{d}) d \mbf{X}_f, \label{eq: continuous barrier}
\end{equation}
where the barrier energy density $b(d, \hat{d})$ is piecewise smooth and only activated when $d < \hat{d}$, improving efficiency and approximately satisfying the complimentarity slackness condition. As $d$ approaches 0, the value of $b(d, \hat{d})$ monotonically increases to infinity, providing arbitrarily large repulsion to avoid interpenetration.

Since our solids and fluids domains are respectively discretized as meshes and particles, the barrier potential (Eq. \ref{eq: continuous barrier}) in 3D can be numerically integrated as
\begin{equation}
\begin{aligned}
    B_\text{sf}(\mbf{x}_s, \mbf{x}_f) &= \sum_{q \in \mathcal{Q}_f} s_q b(\min_{e \in \mathcal{B}_s} d^{PT}(\mbf{x}_q, e), \hat{d}) \\
                        & = \sum_{q \in \mathcal{Q}_f} s_q \max_{e \in \mathcal{B}_s}b(d^{PT}(\mbf{x}_q, e), \hat{d}),
\end{aligned}
\end{equation}
where $\mathcal{Q}_f$ is the set of all SPH fluid particles, $\mathcal{B}_s$ is the set of all boundary triangles of the solids, $s_q = \pi (\frac{3V_0}{4\pi})^{\frac{2}{3}}$ is the integration weight (boundary area) of each fluid particle, and $d_{PT}(\mbf{x}_q, e)$ measures the distance between particle $\mbf{x}_q$ and triangle $e$. Here the min-max transformation is based on the non-ascending property of the barrier function. However, the max operator here makes the barrier potential challenging to be efficiently optimized by gradient-based methods. Fortunately, due to the local support of the barrier function $b(d, \hat{d})$ as $\hat{d}$ is small, we can simply approximate the barrier potential as
\begin{equation}
    B_\text{sf}(\mbf{x}_s, \mbf{x}_f) = \sum_{q \in \mathcal{Q}_f} \sum_{e \in \mathcal{B}_s}s_q b(d^{PT}(\mbf{x}_q, e), \hat{d}),
\end{equation}
which may result in overestimated contact forces near the edges and nodes on the mesh boundary, but we have not observed any artifacts in our experiments.

\subsubsection{Friction Potential}
Following \citet{li2020incremental}, we model the local friction forces $\mbf{f}_k$ for every active solid-fluid contact pair k. Formally, the friction force is defined as
\begin{equation}
    \mbf{f}_k(\mbf{x}_s, \mbf{x}_f) = -\mu_t \lambda_k T_k(\mbf{x}_s, \mbf{x}_f) f_1(\|\mbf{u}_k\|)\frac{\mbf{u}_k}{\|\mbf{u}_k\|},
\end{equation}
where $\lambda_k$ is the contact force magnitude, $T_k(\mbf{x}_s, \mbf{x}_f) \in \mathbb{R}^{3n \times 2}$ is the consistently oriented sliding basis, and $\mbf{u}_k$ is the relative sliding displacement, which can be computed as $\mbf{u}_k = T_k(\mbf{x}_s, \mbf{x}_f)^T ([\mbf{x}_s^T, \mbf{x}_f^T]^T - [\mbf{x}_s^n, \mbf{x}_f^n]^T)$. Here $f_1$ is a smoothly approximated function designed for the smooth transition between sticking and sliding modes. To make this friction formulation fit into optimization time integration, \citet{li2020incremental} further approximated the sliding basis $T(\mbf{x}_s, \mbf{x}_f)$ and contact force $\lambda_k(\mbf{x}_s, \mbf{x}_f)$ explicitly as $T(\mbf{x}^n_s, \mbf{x}^n_f)$ and $\lambda_k(\mbf{x}^n_s, \mbf{x}^n_f)$. Then the semi-implicit friction force is integrable with the friction potential computed as
\begin{equation}
    D_\text{sf}(\mbf{x}_s, \mbf{x}_f) = \sum_{k \in \mathcal{A}^n} \mu_t \lambda^n_k f_0(\|\mbf{u}_k\|),
\end{equation}
where $f_0$ is defined by the relation $f_0' = f_1$ and $\mathcal{A}^n$ is the set containing all activate particle-triangle contact pairs at the previous time step $n$.

\subsection{Optimization Time Integrator}
With the above potential energies modeling all the solid and fluid forces, now we can build a unified two-way solid-fluid coupling framework. By stacking all nodal positions and velocities of SPH particles and FEM nodes as $\mathbf{x} = [\mathbf{x}_f^T, \mathbf{x}_s^T]^T$ and $\mathbf{v} = [\mathbf{v}_f^T, \mathbf{v}_s^T]^T$, we define $\Psi(\mbf{x}) = \Psi_s(\mbf{x}_s)$, $P(\mbf{x}) = P_I(\mbf{x}_f) + P_V(\mbf{x}_f)$ and $C_\text{sf}(\mbf{x}) = B_\text{sf}(\mbf{x}_s, \mbf{x}_f) + D_\text{sf}(\mbf{x}_s, \mbf{x}_f)$. 
Combined with the solid-solid contact potential $C_\text{ss}(\mbf{x})$ from IPC, our solid-fluid coupling problem can be solved in a monolithic manner applying implicit Euler time integration
\begin{equation}
    \begin{cases}
        \ \mathbf{v}^{n+1} =
            \mathbf{v}^{n}
         + h \mathbf{M}^{-1} (\mbf{f}_\text{ext} - \nabla P(\mbf{x}^{n+1})-\nabla \Psi(\mathbf{x}^{n+1}) -\nabla C(\mathbf{x}^{n+1}) 
         ) \\
        
        \ \mathbf{x}^{n+1} = \mathbf{x}^n + h \mathbf{v}^{n+1}
    \end{cases},
    \label{eq:monolithic_discrete_PDE}
\end{equation}
which is equivalent to
\begin{equation}
\label{optimization problem}
    \mathbf{x}^{n+1} = \text{arg}\min_\mathbf{x} \frac{1}{2}\| \mathbf{x} - \hat{\mathbf{x}}^n\|^2_\mathbf{M} + h^2 (P(\mathbf{x}) + \Psi(\mathbf{x}) + C(\mathbf{x}))
\end{equation}
with the mass matrix $\mathbf{M}$, time step size $h$, the predictive position $\hat{\mathbf{x}}^n = \mathbf{x}^n + h \mathbf{v}^n + h^2 \mathbf{M}^{-1} \mathbf{f}_{\text{ext}}$ and the total contact potential $C(\mbf{x}) = C_\text{sf}(\mbf{x}) + C_\text{ss}(\mbf{x})$. 

\begin{figure*}[t]
  \centering
  \includegraphics[width=\linewidth]{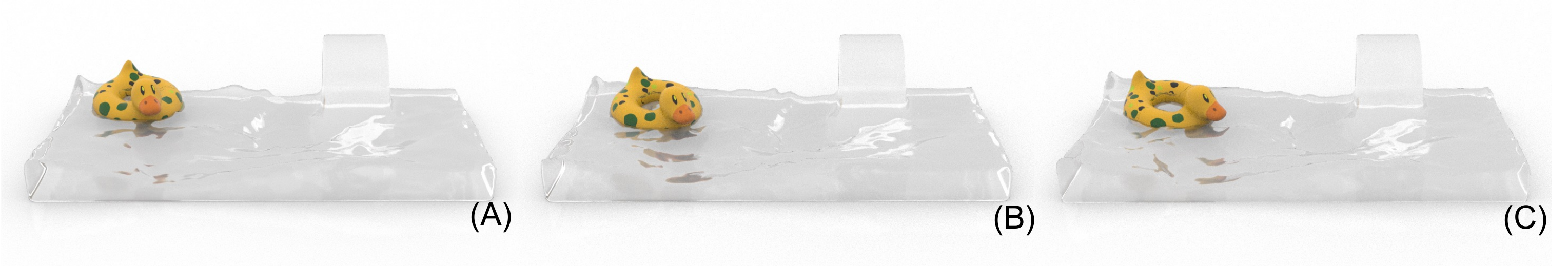}
  \caption{\textbf{Bob} simulated with (A) Joint Optimization, (B) Time Splitting with Contact Proxy, and (C) Baseline Time Splitting. For this example, baseline time splitting can also produce visually plausible results, and our proxy-assisted scheme is $3\times$ faster than joint optimization.}
  \label{fig:bob}
\end{figure*}
\section{Efficient Solver}
A straightforward way to robustly solve the time-stepping optimization problem (Eq.~\ref{optimization problem}) is to apply the projected Newton's method with line search \cite{li2020incremental}. At every iteration, the search direction $\mathbf{p}$ can be computed by solving the linear system
\begin{equation}
    \begin{bmatrix}
        \mathbf{H}_f & \mathbf{G} 
        \\ \mathbf{G}^T & \mathbf{H}_s
    \end{bmatrix} \mathbf{p}= 
    \begin{bmatrix}
        \mathbf{g}_f \\ \mathbf{g}_s
    \end{bmatrix}.
    \label{eq:search_dir_joint}
\end{equation}
Here $\mathbf{H}_f$ and $\mathbf{H}_s$ are the (projected) Hessian matrices w.r.t. the position of fluids and solids respectively, and $\mathbf{G} = \frac{\partial^2 E}{\partial\mathbf{x}_f \partial \mathbf{x}_s}$ denotes the coupling submatrix. Nevertheless, solving this linear system can be a severe bottleneck in practice. One reason is that SPH techniques need sufficient neighbors to accurately approximate physical quantities, which results in a much larger and denser fluid Hessian matrix $\mathbf{H}_f$ compared to the solid one. In addition, the optimization may require many iterations to converge due to the sharpness of barrier energy, especially in contact-rich cases. 

Since our fluid energies are all quadratic, we separate them from the highly nonlinear solids and contact energies via a robust time splitting scheme (\S~\ref{sec:time_splitting}) so that the fluid part can be solved within a single Newton iteration per time step. We then propose efficient methods to solve the domain-decomposed linear systems (\S~\ref{sec:linear_solve}).

\subsection{Time Splitting}\label{sec:time_splitting}

\subsubsection{Baseline Time Splitting}

Intuitively, we can split the original time integration into a \textbf{fluid phase}
\begin{equation}
    \begin{cases}
        \ \mathbf{v}_f^{n+1/2} = \mathbf{v}_f^n + h \mathbf{M}_f^{-1} (-\nabla_f P([(\mathbf{x}_f^{n+1/2})^T,(\mathbf{x}_s^n)^T]^T) + \mathbf{f}_f)\\

        \ \mathbf{x}_f^{n+1/2} = \mathbf{x}_f^n + h \mathbf{v}_f^{n+1/2}
    \end{cases}
\end{equation}
and a \textbf{solid-coupling phase}
\begin{equation}
    \begin{cases}
        \ \mathbf{v}^{n+1} = 
        \begin{bmatrix}
            \mathbf{v}_f^{n+1/2} \\
            \mathbf{v}_s^{n}
        \end{bmatrix} + h \mathbf{M}^{-1} (-\nabla \Psi(\mathbf{x}^{n+1}) -\nabla C(\mathbf{x}^{n+1}) + 
        \begin{bmatrix}
            \mathbf{0} \\
            \mathbf{f}_s
        \end{bmatrix})\\
        
        \ \mathbf{x}^{n+1} = \mathbf{x}^n + h \mathbf{v}^{n+1}
    \end{cases},
\end{equation}
where $\mathbf{f}_f$ and $\mathbf{f}_s$ are the external forces on the fluids and the solids respectively. 
In the fluid phase, we solve for an intermediate state for the fluid particles in a single Newton's iteration, ignoring contact.
Then the highly nonlinear barrier force is resolved in the solid-coupling phase along with elasticity, where the fluid Hessian $\mathbf{H}_f$ reduces to a block-diagonal matrix $\frac{\partial^2 C(\mathbf{x})}{\partial \mathbf{x}_f^2}$. In this setting, nonlinear optimization only happens for fluid boundaries and solid DOFs in the solid-coupling phase. The details of this \textit{Baseline Time Splitting Scheme} can be found in the appendix. 

Although this baseline splitting strategy indeed brings a significant performance gain, severe instabilities can happen at the solid-fluid interface if the time step size is not sufficiently small, especially when simulating viscous fluids (Fig.~\ref{fig: contact proxy}).
For example, fluid particles may stick to the solid boundaries. This is an artifact also seen in existing SPH fluid solvers, and it is typically addressed by sampling particles at solid boundaries to exert boundary pressures \cite{becker2009direct, ihmsen2010boundary, akinci2012versatile}. 
In light of this, we consistently augment the fluid phase with proxy forces for solid-fluid contact to improve stability while avoiding any particle sampling overhead.
\subsubsection{Time Splitting with Contact Proxy}\label{sec:TSCP}
We introduce a solid-fluid contact proxy energy $\hcsf (\mathbf{x})$ into the fluid phase to efficiently exert approximated interaction forces between the boundaries of solids and fluids. In the following discussions, we will also write contact energy $C(\mathbf{x})$ as the sum of the solid-fluid part ($\csf (\mathbf{x})$) and the solid-solid part ($\css (\mathbf{x})$) for clarity.
To ensure consistency with the original PDE, we cancel the contribution of this contact proxy in the solid-coupling phase. 
The resulting time integration becomes
\begin{equation}
    \begin{aligned}
    & \begin{cases}
        \ \mathbf{v}^{n+1/2} = \mathbf{v}^n + h \mathbf{M}^{-1} (-\nabla P(\mathbf{x}^{n+1/2}) + \mathbf{f}_\text{ext}
        - \nabla \hat{C}_\text{sf}(\mathbf{x}^{n+1/2}))
        \\
        
        \ \mathbf{x}^{n+1/2} = \mathbf{x}^n + h \mathbf{v}^{n+1/2}
    \end{cases} \\
    & \begin{cases}
        \ \mathbf{v}^{n+1} =
            \mathbf{v}^{n+1/2}
         + h \mathbf{M}^{-1} (-\nabla \Psi(\mathbf{x}^{n+1}) -\nabla C(\mathbf{x}^{n+1}) 
         + \nabla \hat{C}_\text{sf}(\mathbf{x}^{n+1})) \\
        
        \ \mathbf{x}^{n+1} = \mathbf{x}^n + h \mathbf{v}^{n+1}
    \end{cases}
    \end{aligned}
    \label{eq:two_phase_TS}
\end{equation}
where the fluid phase now also implicitly updates the solid boundary nodes near the fluids to an intermediate state and explicitly update all other solid nodes.

For $\hcsf(\mathbf{x})$, a straightforward choice is $\hcsf(\mathbf{x}) = \frac{1}{2}\csf(\mathbf{x})$. But to ensure our fluid phase still only contains linear forces, we apply the 2nd-order Taylor expansion of $\frac{1}{2}\csf(\mathbf{x})$ at $\mathbf{x}^n$ for the approximation in the fluid phase, i.e.
\begin{equation}
    \begin{aligned}
        \hcsf (\mathbf{x}) = \frac{1}{2} \left(\csf (\mathbf{x}^n) + \nabla \csf (\mathbf{x}^n)(\mathbf{x} - \mathbf{x}^n) + \frac{1}{2}\|\mathbf{x} - \mathbf{x}^n\|^2_{\nabla^2  \csf (\mathbf{x}^n)} \right)
    \end{aligned},
\end{equation}
while in the solid-coupling phase, we simply use $\hcsf(\mathbf{x}) = \frac{1}{2}\csf(\mathbf{x})$. In the appendix, we prove that our time splitting scheme with contact proxy only has an $\mathcal{O}(h^4)$ mismatch compared to implicit Euler solution. Reformulating both phases (Eq.~\ref{eq:two_phase_TS}) as optimization problems, we obtain
\begin{equation}
\boxed{
    \begin{aligned}
        \mathbf{x}^{n+1/2} &= \argmin_\mathbf{x} \frac{1}{2}\| \mathbf{x} - \hat{\mathbf{x}}^n\|^2_\mathbf{M} + h^2 (P(\mathbf{x}) + \hcsf (\mathbf{x})),\\
        \mathbf{x}^{n+1} &= \argmin_\mathbf{x} \frac{1}{2}\|\mathbf{x} - \mathbf{x}^{n+1/2}\|^2_\mathbf{M} + h^2 (\Psi(\mathbf{x}) + \frac{1}{2}\csf (\mathbf{x}) + \css (\mathbf{x})),
    \end{aligned}
}
\end{equation}
where $\hat{\mathbf{x}}^n = \mathbf{x}^n + h \mathbf{v}^n + h^2 \mathbf{M}^{-1} \mathbf{f}_\text{ext}$ with $\mathbf{f}_\text{ext} = [\mathbf{f}_f^T, \mathbf{f}_s^T]^T$.

In addition to avoiding fluid particle sticking issues without extra expensive costs, another benefit of our method is that it helps reduce the number of Newton's iterations for solving the problem. Typically, the barrier method takes many Newton iterations when resolving high-speed impacts. With our scheme, when high-speed fluid particles are colliding with a deformable object, their speed will be significantly reduced after the fluid phase due to the contact proxy. The reduced speed will then be taken into the solid-coupling phase, which makes the nonlinear optimization easier to solve (by having less contact constraint set changes). The details of our proxy-based time splitting scheme can be found in Alg.~\ref{alg: time splitting with proxies}. 
\begin{algorithm}
\caption{Time Splitting with Contact Proxy}\label{alg: time splitting with proxies}
\begin{algorithmic} [1]
\State $\mathbf{x} \gets \mathbf{x}^n$, \ $\hat{\mathbf{x}}^n \gets \mathbf{x}^n + h \mathbf{v}^n + h^2 \mathbf{M}^{-1} \mathbf{f}_\text{ext}$
\State $\text{SPH Neighbor Search } \& \text{ Density Update}$
\State $\hcsf (\mathbf{x}) \gets \text{2nd Taylor Expansion of } \frac{1}{2} \csf (\mathbf{x}) \text{ at } \mathbf{x}=\mathbf{x}^n$
\State // Fluid Phase
\State $\mathbf{H} \gets h^2 \left(\nabla^2 P(\mathbf{x}) + \nabla^2 \hcsf (\mathbf{x})\right) + \mathbf{M}$
\State $\mathbf{p} \gets -\mathbf{H}^{-1}\left(h^2 (\nabla P(\mathbf{x}) + \nabla \hcsf (\mathbf{x})) + \mathbf{M}(\mathbf{x} - \hat{\mathbf{x}}^n)\right) $
\State $\mathbf{x} \gets \mathbf{x} + \mathbf{p}$
\State $\mathbf{x}^{n+1/2} \gets \mathbf{x}$
\State // Solid-Coupling Phase
\Do
    \State $\mathbf{H} \gets h^2 \left(\nabla^2 \Psi(\mathbf{x}) + \frac{1}{2}\nabla^2 \csf(\mathbf{x}) + \nabla^2 \css(\mathbf{x}) \right) + \mathbf{M}$
    \State $\mathbf{g} \gets h^2 \left(\nabla \Psi(\mathbf{x}) + \frac{1}{2} \nabla \csf (\mathbf{x}) + \nabla \css (\mathbf{x})\right) + \mathbf{M}(\mathbf{x} - \mathbf{x}^{n+1/2})$
    \State $\mathbf{p} \gets -\mathbf{H}^{-1}\mathbf{g} $
    \State $\alpha \gets \text{Backtracking Line Search with CCD}$
    \State $\mathbf{x} \gets \mathbf{x} + \alpha \mathbf{p}$
\doWhile{$\frac{1}{h} \|\mathbf{p}\| > \epsilon$}
\State $\mathbf{x}^{n+1} \gets \mathbf{x}$, \ $\mathbf{v}^{n+1} \gets (\mathbf{x} - \mathbf{x}^n)/h$ \\
\Return $\mathbf{x}^{n+1}, \mathbf{v}^{n+1}$
\end{algorithmic}
\end{algorithm}

Similarly, one can also separate elasticity from contact energy using the contact proxy. In this fashion, we would have a three-phase (fluid, solid, and contact) time splitting scheme
\begin{equation}
\boxed{
    \begin{aligned}
        \mathbf{x}^{n+1/3} &= \argmin_\mathbf{x} \frac{1}{2}\| \mathbf{x} - \hat{\mathbf{x}}^n\|^2_\mathbf{M} + h^2 (P(\mathbf{x}) + \hcsf (\mathbf{x})),\\
        \mathbf{x}^{n+2/3} &= \argmin_\mathbf{x} \frac{1}{2}\|\mathbf{x} - \mathbf{x}^{n+1/3}\|^2_\mathbf{M} + h^2 (\Psi(\mathbf{x}) + \hcsf (\mathbf{x}) + \hcss(\mathbf{x})),\\
        \mathbf{x}^{n+1} &= \argmin_\mathbf{x} \frac{1}{2}\|\mathbf{x} - \mathbf{x}^{n+2/3}\|^2_\mathbf{M} + h^2 (\frac{1}{3}\csf (\mathbf{x}) + \frac{1}{2}\css (\mathbf{x})),
    \end{aligned}
}
    \label{eq: three phase}
\end{equation}
where $\hcsf(\mathbf{x})$ and $\hcss(\mathbf{x})$ are the 2nd-order Taylor expansion of $\frac{1}{3}\csf(\mathbf{x})$ and $\frac{1}{2}\css(\mathbf{x})$ respectively. However, this aggressive splitting scheme only applies to inversion-robust constitutive models, e.g. the fixed corotated model\ \cite{stomakhin2012energetically}. While inversion can be prevented with guarantee at the solid phase where the elasticity energy is considered, it may not hold at the contact phase. Despite this limitation, the three-phase splitting scheme can still work properly for inversion-robust constitutive models in practice to further accelerate the simulation.


\subsection{Solving Linear Systems} \label{sec:linear_solve}
In our time splitting scheme, solving large sparse linear systems dominates both the computational and memory costs of each phase. We thus devise matrix-free and Schur-complement based strategies to solve them efficiently.

\subsubsection{Fluid Phase}\label{sec:mtr_free_cg}
Since 2-ring neighbors of SPH particles need to be considered in our formulation, both constructing and directly factorizing the Hessian matrix can cost a significant amount of time and memory. Therefore, we devise a matrix-free conjugate gradient (CG) solver to efficiently solve for the intermediate state of fluids.

As all energy potentials are quadratic in this phase, the energy gradient $\mathbf{g}(\mathbf{x})$ is merely a linear function of $\mathbf{x}$ with constant coefficient matrix $\mathbf{H}(\mathbf{x})$. Thus, the product between $\mathbf{H}(\mathbf{x})$ and an arbitrary vector $\mathbf{p}$ can be expressed as
\begin{equation}
\label{eq: matrix free}
    \mathbf{H}(\mathbf{x}) \mathbf{p} = \mathbf{g}(\mathbf{p}) - \mathbf{g}(\mathbf{0}).
\end{equation}
This allows us to compute gradients to evaluate the matrix-vector product, and we only need to acquire the $3 \times 3$ diagonal blocks of the Hessian for block-Jacobi preconditioning in our CG solver. 

\subsubsection{Solid-Coupling Phase}
As the fluid energy potential is not included in this phase, the components of the Hessian matrix become
\begin{equation}
    \begin{aligned}
        \mathbf{H}_f = \frac{\partial^2 C(\mathbf{x})}{\partial \mathbf{x}_f^2}, \quad 
        \mathbf{G} = \frac{\partial^2 C(\mathbf{x})}{\partial \mathbf{x}_s \partial \mathbf{x}_f}, \quad
        \mathbf{H}_s = \frac{\partial^2 C(\mathbf{x})}{\partial \mathbf{x}_s^2} + \frac{\partial^2 \Psi(\mathbf{x})}{\partial \mathbf{x}_s^2}.
    \end{aligned}
\end{equation}
Although this linear system is no longer that intractable, it is not optimal to directly factorize the whole system given the considerable amount of nonzeros in $\mathbf{H}_f$ and $\mathbf{G}$ when fluid resolution is high. 

We thus design a domain decomposed linear solver that treats $\mathbf{H}_f$ and $\mathbf{H}_s$ separately. Based on Schur complement\ \cite{zhang2006schur}, the inverse of our Hessian matrix can be expressed as
\begin{equation}
   \mathbf{H}^{-1} =  \begin{bmatrix}
                        \mathbf{H}_f^{-1} + \mathbf{H}_f^{-1}\mathbf{G} (\mathbf{H}/\mathbf{H}_f)^{-1} \mathbf{G}^T \mathbf{H}_f & -\mathbf{H}_f^{-1}\mathbf{G} (\mathbf{H}/\mathbf{H}_f)^{-1} 
                        \\ -(\mathbf{H}/\mathbf{H}_f)^{-1} \mathbf{G}^T \mathbf{H}_f & (\mathbf{H}/\mathbf{H}_f)^{-1}
                    \end{bmatrix},
\end{equation}
where $\mathbf{H}/\mathbf{H}_f = \mathbf{H}_s - \mathbf{G}^T \mathbf{H}_f^{-1} \mathbf{G}$ is the Schur complement of block $\mathbf{H}_f$.
Since the nonzeros of $\mathbf{H}_f$ only exist in the diagonal blocks, it is trivial to obtain its inverse matrix $\mathbf{H}_f^{-1}$. We can then apply the CHOLMOD\ \cite{chen2008algorithm} LLT solver to factorize $\mathbf{H}/\mathbf{H}_f$, which is only in the size of solid DOFs, and then the search direction can be computed via matrix-vector products and back-solves.
When there is no solid-fluid interaction, $\mathbf{H}/\mathbf{H}_f$'s sparsity pattern remains identical with $\mathbf{H}_s$. Only when two solid nodes $i$ and $j$ are interacting with the same fluid particle, the $3\times 3$ block $(\mathbf{H}/\mathbf{H}_f)_{i,j}$ (in 3D) will become non-zero. Typically, this only happens for neighboring mesh primitives and thus the sparsity pattern of $\mathbf{H}/\mathbf{H}_f$ is mostly nice. 

Note that when the three-phase time splitting scheme (Eq.~\ref{eq: three phase}) is used, our domain decomposed solver can also be applied to the solid and contact phases since their systems share a similar structure with the solid-coupling phase here.

\section{Experiments and Evaluation}
\begin{figure*}[t]
  \centering
  \includegraphics[width=\linewidth]{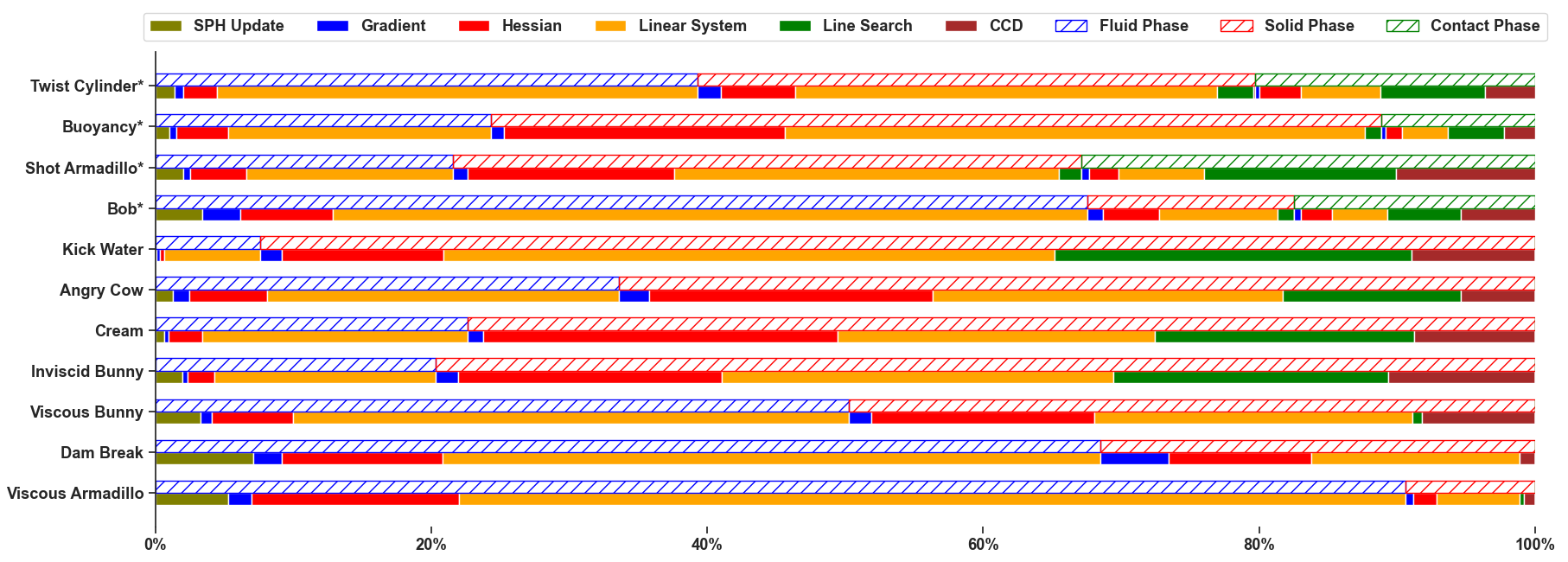}
  \caption{\textbf{Timing breakdown}. We show the timing profile of different simulation phases and plot the proportions of the major routines. Examples marked with * are simulated using our three-phase time splitting scheme. Other examples are generated with the two-phase scheme. 
  In particular, SPH update (including neighborhood search and density update) only occurs in the fluid phase, line search happens in the solid and contact phases for non-linear optimization, and continuous collision detection (CCD) is counted when IPC contact energy is considered.}
  \label{fig: timing breakdown}
\end{figure*}

\begin{table*}
  \begin{tabular}{lcccccccccccc}
    \toprule
    Scene & $\Delta t_{\text{frame}}$ &$\Delta t$ & $N_{\text{fluid}}$ & $N_{\text{solid}}$ &$k_I$ & $\nu_f$ & $d$ & $\rho_f$ & E & $\nu_s$ & $\rho_s$ & $T$\\
    \midrule
    Fig. \ref{fig:bob} Bob &1/24&$4 \times 10^{-3}$ &97K & 2.3K& $2 \times 10^5$ & 0 & $15$ & $1000$ & $1 \times 10^5$ & 0.3 & $500$ & 0.3\\
    Fig. \ref{fig: viscous_armadillo} Viscous Armadillo&1/48&$4 \times 10^{-3}$ &238K & 0& $1 \times 10^5$ & 100 & $10$ & $1200$ & - & - & - & 0.4\\
    Fig. \ref{fig: shot armadillo} Shot Armadillo&1/24& $4 \times 10^{-3}$ &103K & 16K& $1 \times 10^5$ & 0 & $10$ & $1000$ & $1 \times 10^5$ & 0.3 & $200$ & 1.3\\
    Fig. \ref{figure: fluid dynamics comparison} Dam Break&1/24& $5 \times 10^{-3}$ &280K & 0& $2 \times 10^5$ & 0.005 & $25$ & $1000$ & - & - & - & 0.4\\
    Fig. \ref{fig: inviscid bunny} Liquid Bunnys &1/50& $4 \times 10^{-3}$ &52K & 3.7K& $1 \times 10^5$ & 0 & $10$ & $1000$ & $4 \times 10^3$ & 0.49 & $200$ & 0.4\\
    Fig. \ref{fig: liquid bunny2} Liquid Bunnys &1/50& $4 \times 10^{-3}$ &101K & 4.5K& $6 \times 10^4$ & 0 & $6.4$ & $1000$ & $1 \times 10^3$ & 0.49 & 200 & 1.0 \\
    Fig. \ref{fig: buoyancy} Buoyancy&1/24& $5 \times 10^{-3}$ &787K & 66K& $2 \times 10^5$ & 1 & $10$ & $1000$ & $1 \times 10^5$ & 0.4 & $200/700/1200$ & 5.9\\
    Fig. \ref{fig: twist cylinder} Twist Cylinder*&1/24& $5 \times 10^{-3}$ &486K & 12K& $4 \times 10^4$ & 0 & $5$ & $1000$ & - & - & 500 & 7.9\\
    Fig. \ref{fig: cream} Cream &1/24& $4 \times 10^{-3}$ &159K & 9K& $3 \times 10^4$ & 25 & $3$ & $1000$ & $5 \times 10^8$ & 0.49 & $1000$ & 1.8\\
    Fig. \ref{fig: angry cow} Angry Cow*&1/24& $5 \times 10^{-3}$ &789K & 13K& $1 \times 10^5$ & 0.2 & $10$ & $1000$ & $1 \times 10^5$ & 0.45 & $100/700$ & 4.9\\
    Fig. \ref{fig: kick} Kick Water* &1/24& $6 \times 10^{-3}$ &1M & 43K& $2.5 \times 10^5$ & 0.1 & $25$ & $1000$ & - & - & 500 & 37.9\\
  \bottomrule
  \end{tabular}
\caption{\textbf{Simulation statistics} including duration of each frame ($\Delta t_{\text{frame}}, [s]$), time step size upperbound ($\Delta t, [s]$), number of fluid particles ($N_{\text{fluid}}$), number of solid vertices ($N_{\text{solid}}$), incompressibility coefficient ($k_I, [Pa]$), dynamic viscosity ($\nu_f, [Pa\cdot s]$), fluid particle diameter ($d, [mm]$), fluid density ($\rho_f, [kg/m^3]$), Young's modulus ($E, [Pa]$), Possion's ratio ($\nu_s$), solid density ($\rho_s, [kg/m^3]$) and the average simulation time for each frame ($T, [min]$). Timing statistics are measured on a 24-core 3.50GHz Intel i9-10920X machine except for Fig. \ref{fig: ElastoMonolith comparison}, which is tested on the “e2-standard-8” (8 cores with 32GB RAM) Google Compute Engine. Note that examples marked with * contain codimensional materials, whose parameter settings are not covered here.}
\label{tab:experiment parameters}
\end{table*}
Our code is implemented in C++ with Eigen for basic linear algebra operations and Intel TBB for multi-threading. The time step size of all our simulations is adaptively chosen by the SPH CFL condition and a user-defined upper bound. We set the support radius of our SPH kernel function to $2d$, where $d$ is the particle diameter. In our implementation, we use the cubic Spline kernel for density estimation and the Spiky kernel for gradient calculation. For Fig. \ref{fig: twist cylinder}, \ref{fig: buoyancy}, \ref{fig: shot armadillo} and \ref{fig:bob}, we employ our three-phase time splitting scheme, showing its efficacy when the constitutive models are compatible with mesh inversion. For the rest of the simulations, we stick with our two-phase time splitting scheme. Most experiments are performed on a 24-core 3.50GHz Intel i9-10920X machine, except for the comparative study with ElastoMonolith\ \cite{takahashi2022elastomonolith}. We demonstrate that our method achieves efficient and robust solid-fluid coupling. The parameters and timing breakdown of all simulations are provided in Table~\ref{tab:experiment parameters} and Fig.~\ref{fig: timing breakdown} respectively.

\begin{figure}[t]
    \centering
    \begin{subfigure}[b]{0.47\textwidth}
         \centering
         \includegraphics[width=\textwidth]{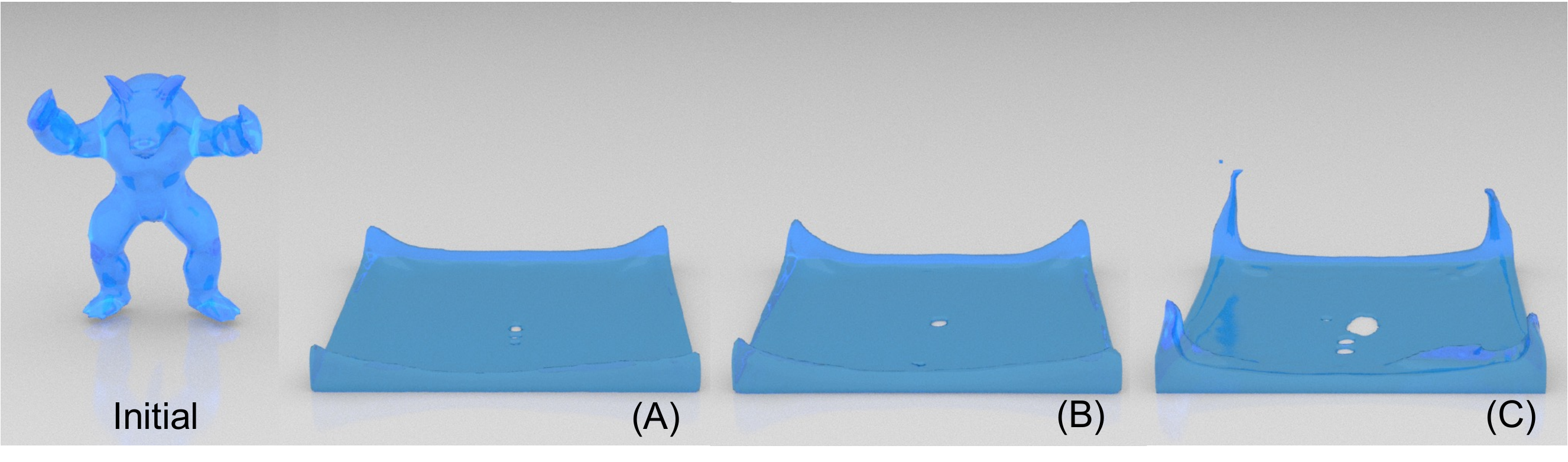}
         \vspace{-0.5cm}
         \caption{A \textbf{viscous armadillo} dropped onto the ground.}
         \label{fig: viscous_armadillo}
         \vspace{0.2cm}
    \end{subfigure}
    \hfill
    \begin{subfigure}[b]{0.47\textwidth}
         \centering
         \includegraphics[width=\textwidth]{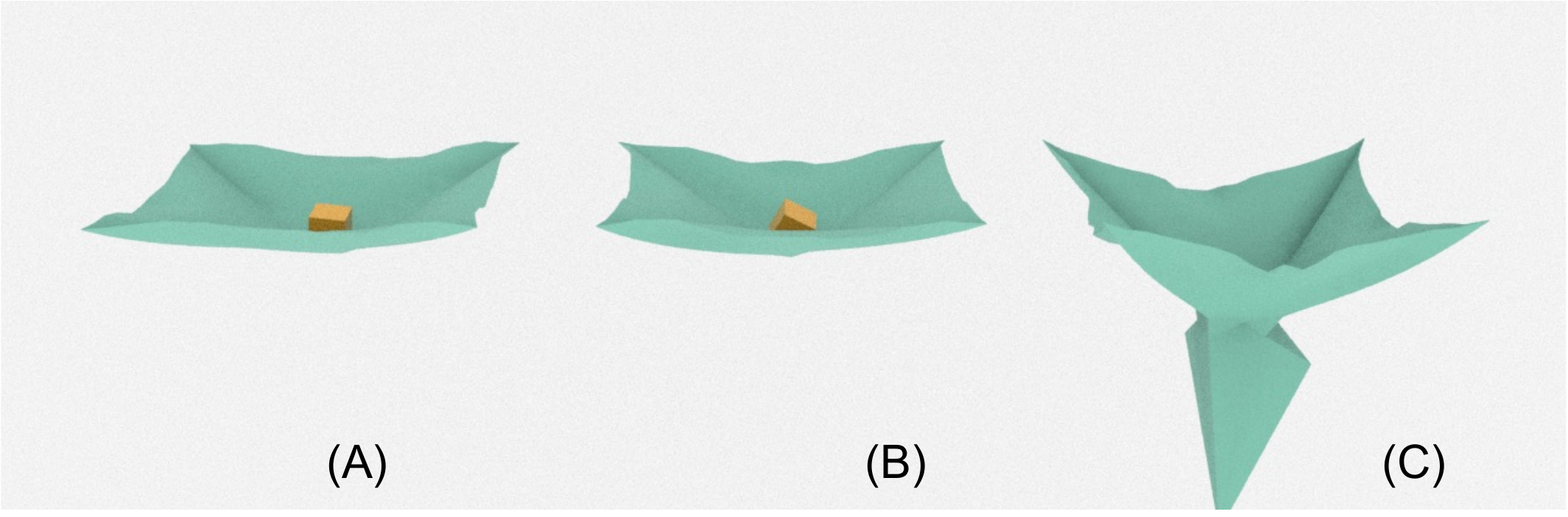}
         \vspace{-0.5cm}
         \caption{\textbf{Cube on cloth}. An elastic cube is dropped onto a square cloth with four corners fixed.}
         \label{fig: cube on cloth}
    \end{subfigure}
    \vspace{-0.2cm}
  \caption{Simulation results of (A) Joint Optimization, (B) Time Splitting with Contact Proxy, and (C) Baseline Time Splitting. While directly applying time splitting results in instability at the boundaries, our results with contact proxy are consistent with joint optimization.}
  \label{fig: contact proxy}
\end{figure}

\subsection{Ablation Study}
\subsubsection{Time Splitting Evaluation}
\begin{table}
  \label{tab:scheme comparison}
  \begin{tabular}{lccccc}
    \toprule
    Scene  & Scheme & Sec/Frame & \# Newton Iter./Frame\\
    \midrule
    Fig. \ref{fig:bob} & Joint/TS/TSCP & 66.1 / 38.0 / 22.5 & 63.5 / 117.3 / 37.1 \\
    Fig. \ref{fig: viscous_armadillo} & Joint/TS/TSCP & 41.3 / 32.3 / 25.5 & 16.5 / 29.0 / 10.5\\
  \bottomrule
  \end{tabular}
\caption{\textbf{Statistics of different time stepping schemes}: Joint Optimization (Joint), Baseline Time Splitting (TS) and Time Splitting with Contact Proxy (TSCP). Our proposed TSCP is much faster than both the Joint and TS.}
\label{tab: time splittimg comparison}
\end{table}

Three simulations (Fig. \ref{fig:bob} and Fig. \ref{fig: contact proxy}) are performed to demonstrate the efficiency of time splitting and the efficacy of our proposed contact proxy on maintaining stability.

To begin with, we need to take care of choosing a proper time step $h$. First of all, it has to be restricted by the CFL condition. Otherwise, severe volume loss may be observed due to SPH approximation error. Additionally, in contrast to the joint optimization (Eq.~\ref{optimization problem}), the time splitting scheme usually requires smaller time steps to stay stable, which imposes a second time step constraint. However, we observed that in practice, even using the largest CFL time step, our proxy-assisted time splitting can still work properly and produce stable simulation results. Hence, for comparison, we use the largest CFL time step for both schemes to maximize their performance as smaller $h$ typically takes more Newton's iterations in total to simulate a frame. 
For joint optimization, since direct factorization is intractable, we solve Eq.~\ref{eq:search_dir_joint} using the block-Jacobi preconditioned conjugate gradient solver with the fluid part matrix free.

As shown in Table \ref{tab: time splittimg comparison}, even in these simple examples, our time splitting scheme is significantly (up to $3\times$) faster than joint optimization, especially for cases (e.g. Fig.~\ref{fig:bob}) involving contacts between fluids and deformable solids. This improvement stems from no longer having to solve for incompressibility of fluids repeatedly within a time step. 
Moreover, one can also find out that Newton's iterations are much less with our proxy-assisted time splitting scheme. 
As discussed in \S~\ref{sec:TSCP}, this is because the challenging high-speed impacts are already partially resolved in the fluid phase. 
Another benefit of time splitting is the support of different error tolerances for the two phases. Errors in the fluid phase are sourced from the solution deviation of the CG solver, while in the solid phase they are directly controlled by the tolerance of Newton's method. Typically, setting a slightly higher tolerance for fluids yields better performance while still producing visually plausible results.

Aside from efficiency, our proposed contact proxy also improves the stability of time splitting scheme. Though simulation results of the baseline time splitting scheme look fine in the case of inviscid fluids, situations get worse when it is applied to viscous fluids. In Fig. \ref{fig: viscous_armadillo}, a viscous armadillo is dropped to the ground. In this example, the baseline time splitting scheme produces severe sticky artifacts at the boundary, and the fluid surface could not finally calm down. 
By consistently applying our contact proxy to exert boundary pressure in the fluid phase, the artifacts can be well resolved as demonstrated in Fig. \ref{fig: viscous_armadillo}.
Similarly, our idea of contact proxy is also applicable to further separate elasticity from IPC contact while maintaining stability, leading to our three-phase scheme (Fig.~\ref{fig: cube on cloth}).
\begin{figure}[t]
  \centering
  \includegraphics[width=\linewidth]{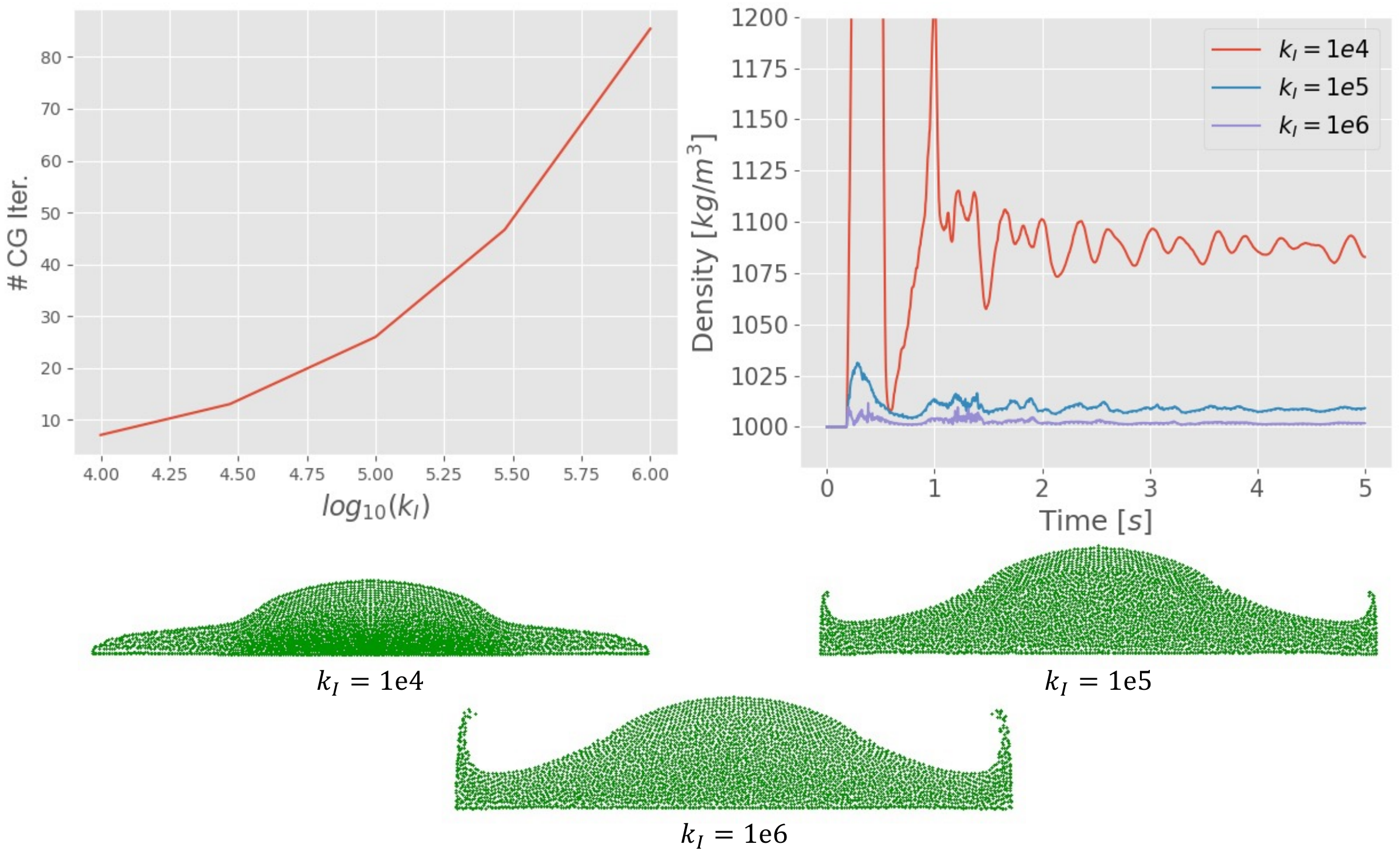}
  \caption{\textbf{Statistics of simulations with different stiffness parameter $k_I$}. A larger $k_I$ preserves volume better but results in more CG iterations. In this case, a proper $k_I = 10^5$ Pa can be set to balance the computational cost and visual artifacts.}
  \label{figure: CG kappa relation}
\end{figure}

\subsubsection{Linear Solver Evaluation}
\begin{table}
  \begin{tabular}{lccccc}
    \toprule
    \multirow{2}{3em}{Solver}  & \multicolumn{2}{c}{Fluid Phase} & Solid Phase & Contact Phase & \multirow{2}{2em}{Mem.}\\
    \cline{2-5}
    & hess & solve & solve & solve & \\
    \midrule
    CG + LLT & 14.9 & 0.49 & 1.45 & 0.43& 12375 \\
    Ours & 0.15 & 0.59 & 1.11 & 0.25 & 1469\\
  \bottomrule
  \end{tabular}
\caption{\textbf{Time and memory cost of different solvers in example \ref{fig: shot armadillo}}. The costs are measured per time step in units [s] and [MB] respectively. The baseline uses the Conjugate Gradient (CG) method for the fluid phase and CHOLMOD LLT for the solid and contact phases. Our method instead employs a matrix-free CG solver for the fluid phase and a domain-decomposed solver for the solid and contact phases, thereby improving efficiency and saving memory.}
\label{tab:solver comparison}
\end{table}
For the fluid phase, we designed a matrix-free conjugate gradient (CG) solver that calculates the matrix-vector product via gradient computation to avoid the expensive computational and memory costs of direct factorization (\S~\ref{sec:mtr_free_cg}).
However, the performance improvement from this approach will be less significant if the number of CG iterations required for convergence is too large, making the cost of computing gradients higher than constructing the Hessian once.
In our fluid phase, the number of CG iterations is proportional to the stiffness $k_I$ of the incompressibility energy. A larger $k_I$ can better preserve the volume of the fluids but also results in a worse-conditioned system, demanding more iterations to converge (Fig. \ref{figure: CG kappa relation}).
In practice, by setting $k_I$ to a proper value, we can efficiently solve the systems within 50 CG iterations without obvious fluid volume loss.


We test the performance of our matrix-free CG solver together with the domain-decomposed solver we designed for the solid-coupling phase on the Shot Armadillo example (Fig.~\ref{fig: shot armadillo}), and present the results in Table~\ref{tab:solver comparison}. 
Our matrix-free CG solver significantly boosts efficiency ($20\times$ faster) and reduces memory costs by avoiding the construction of the Hessian matrix. On the other hand, our domain decomposed solver is 40\% faster than directly factorizing the solid and contact systems.
\subsection{Comparisons}

In this section, we compare our method with several popular SPH fluid solvers and a state-of-the-art solid-fluid coupling method ElastoMonolith \cite{takahashi2022elastomonolith}. We leveraged the open-source library SPlisHSPlasH\footnote{https://github.com/InteractiveComputerGraphics/SPlisHSPlasH} to implement the SPH fluid simulators. To compare our method with ElastoMonolith, we set up two scenes from their paper with identical parameters and run all the simulations using “e2-standard-8” (8 cores with 32GB RAM) Google Compute Engine for fairness.

\begin{figure}[t]
  \centering
  \includegraphics[width=\linewidth]{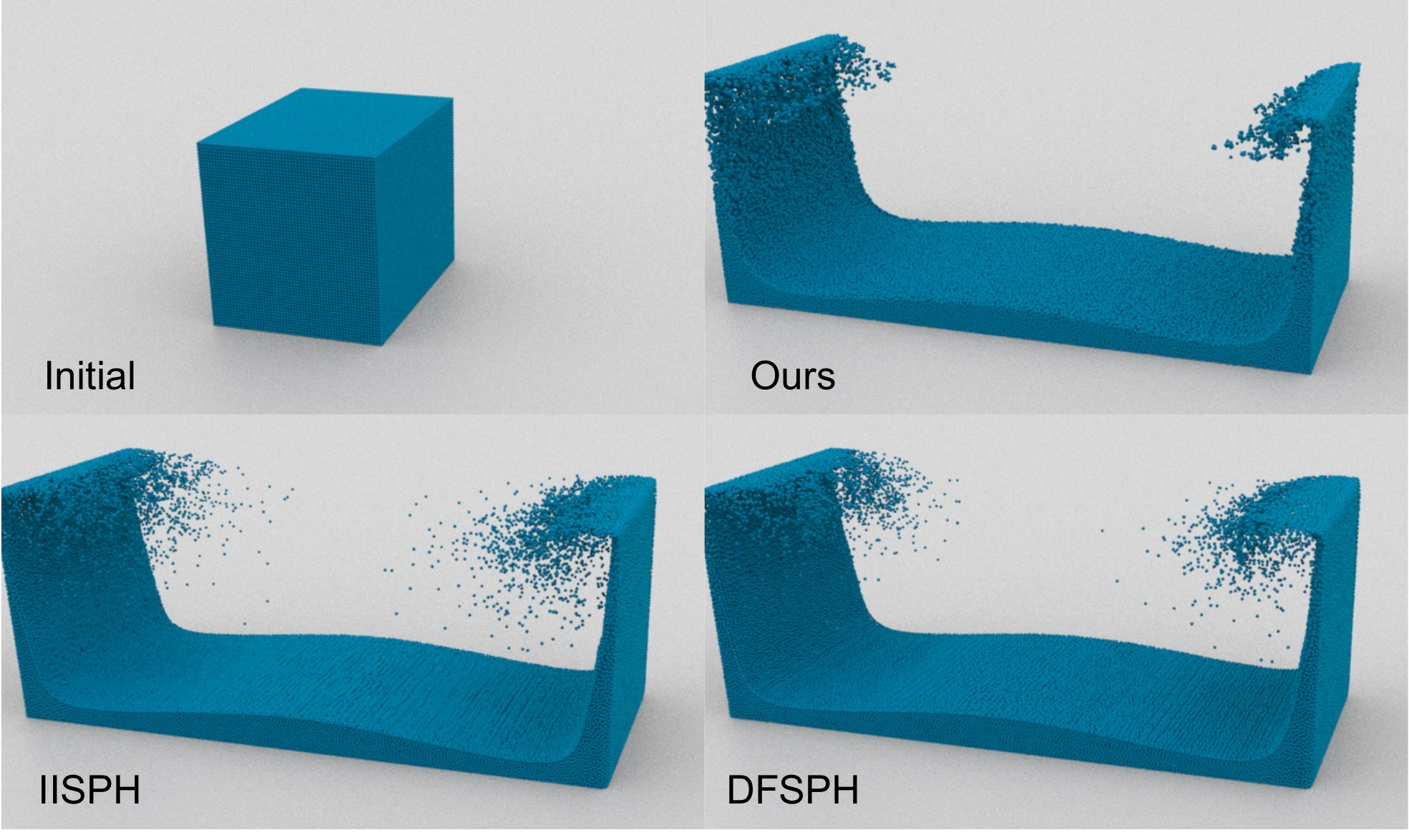}
  \caption{\textbf{Dam break} with 280K SPH particles. Our weakly compressible formulation produces stable fluid dynamics without visually evident volume loss. Compared to incompressible SPH solvers IISPH \cite{ihmsen2013implicit} and DFSPH \cite{bender2015divergence}, our simulation results demonstrate more smooth particle distribution.}
  \label{figure: fluid dynamics comparison}
\end{figure}

\begin{figure}[t]
    \centering
    \begin{subfigure}[b]{0.47\textwidth}
         \centering
         \includegraphics[width=\textwidth]{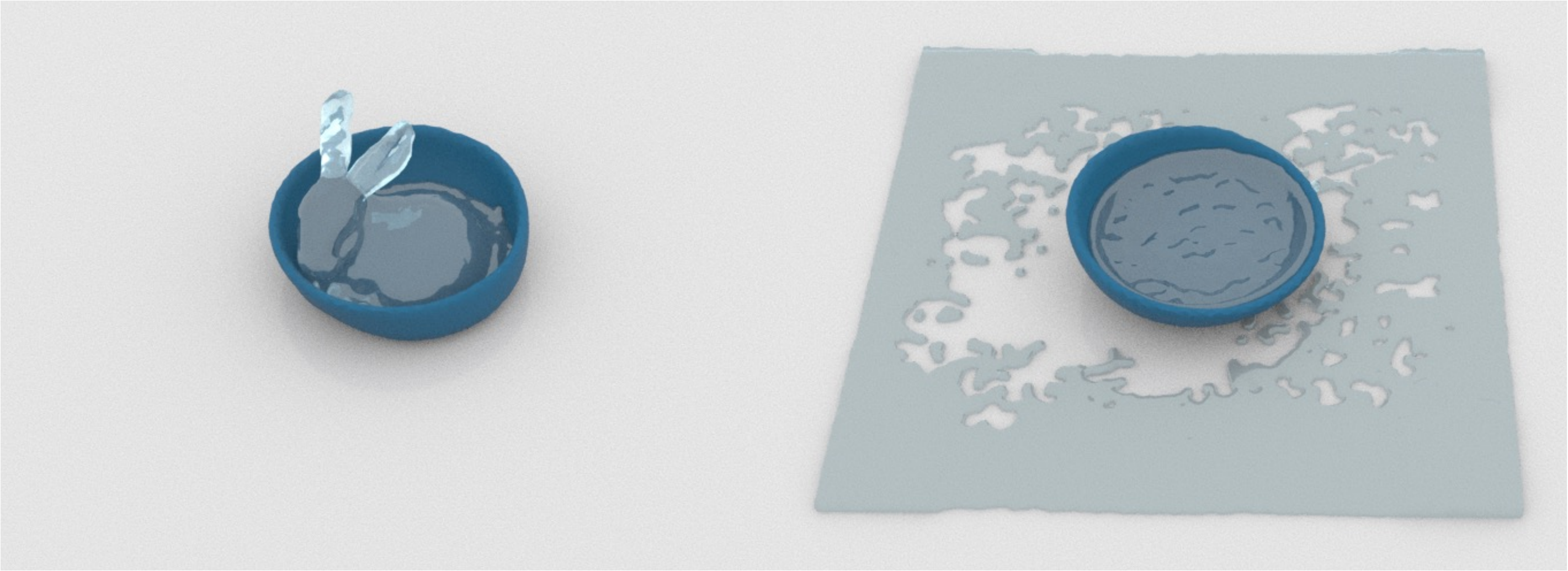}
         \vspace{-0.5cm}
         \caption{A liquid bunny dropped into a bowl.}
         \vspace{0.2cm}
         \label{fig: inviscid bunny}
    \end{subfigure}
    \hfill
    \begin{subfigure}[b]{0.47\textwidth}
         \centering
         \includegraphics[width=\textwidth]{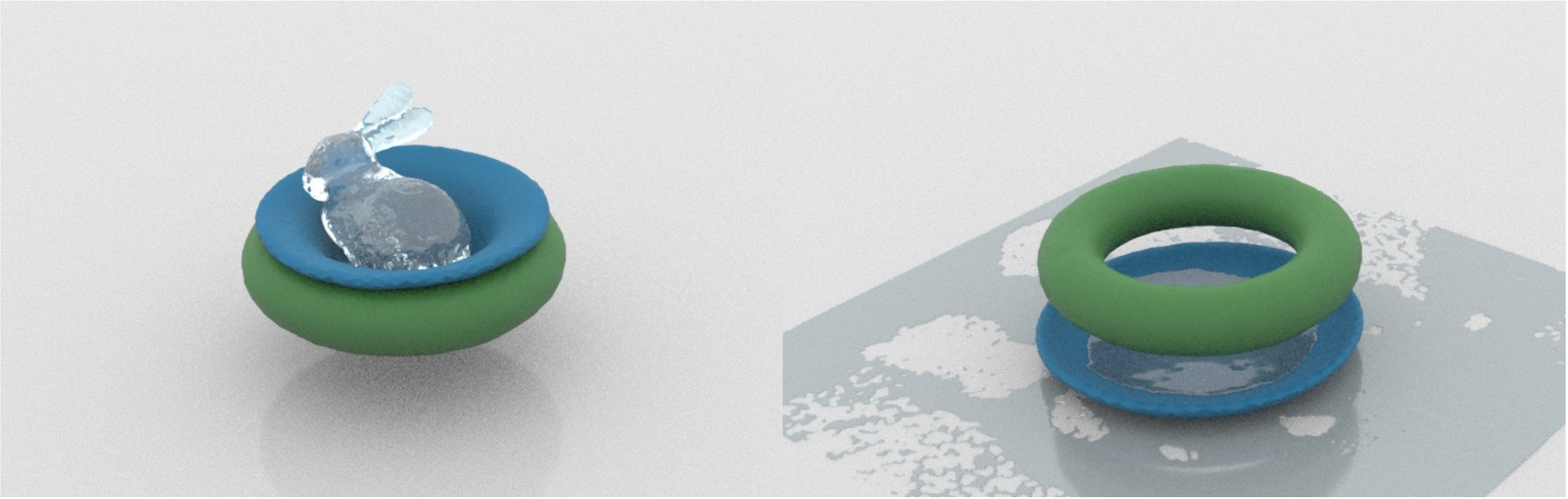}
         \caption{A liquid bunny and an elastic bowl dropped onto a static torus.}
         \vspace{-0.1cm}
         \label{fig: liquid bunny2}
    \end{subfigure}
  \caption{\textbf{Liquid Bunnys}. Compared to ElastoMonolith\ \cite{takahashi2022elastomonolith}, our method achives an over $5\times$ speedup for both of these two examples with exactly the same scene setups.}
  \label{fig: ElastoMonolith comparison}
\end{figure}

\begin{figure*}[t]
  \centering
  \includegraphics[width=\linewidth]{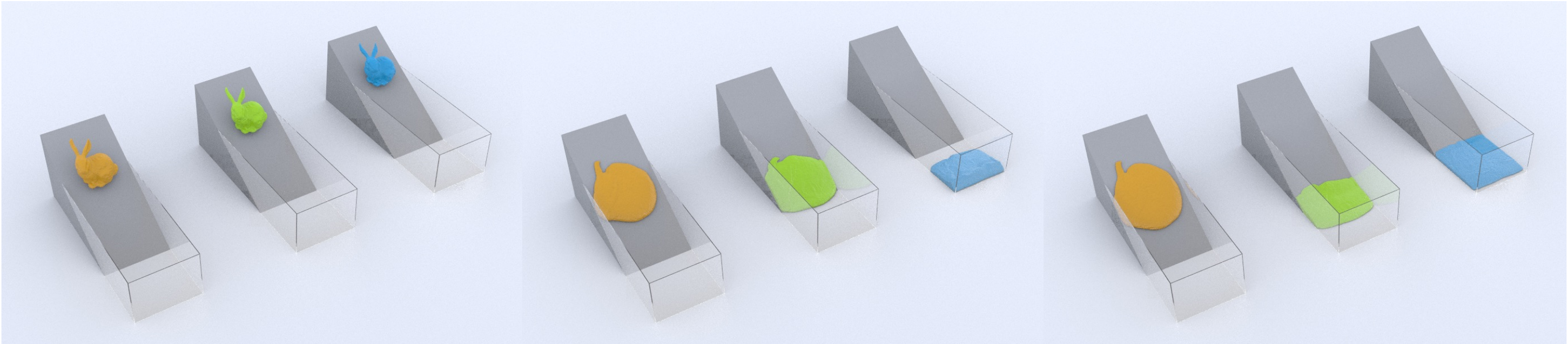}
  \caption{\textbf{Varying friction}. Three viscous bunnies are dropped onto the slope with different coefficients of friction $\mu$ (from left to right: 0.5, 0.03, 0). Our method supports adjustable solid-fluid boundary friction.}
  \label{fig: varying friction}
\end{figure*}

\begin{figure*}[t]
  \centering
  \includegraphics[width=\linewidth]{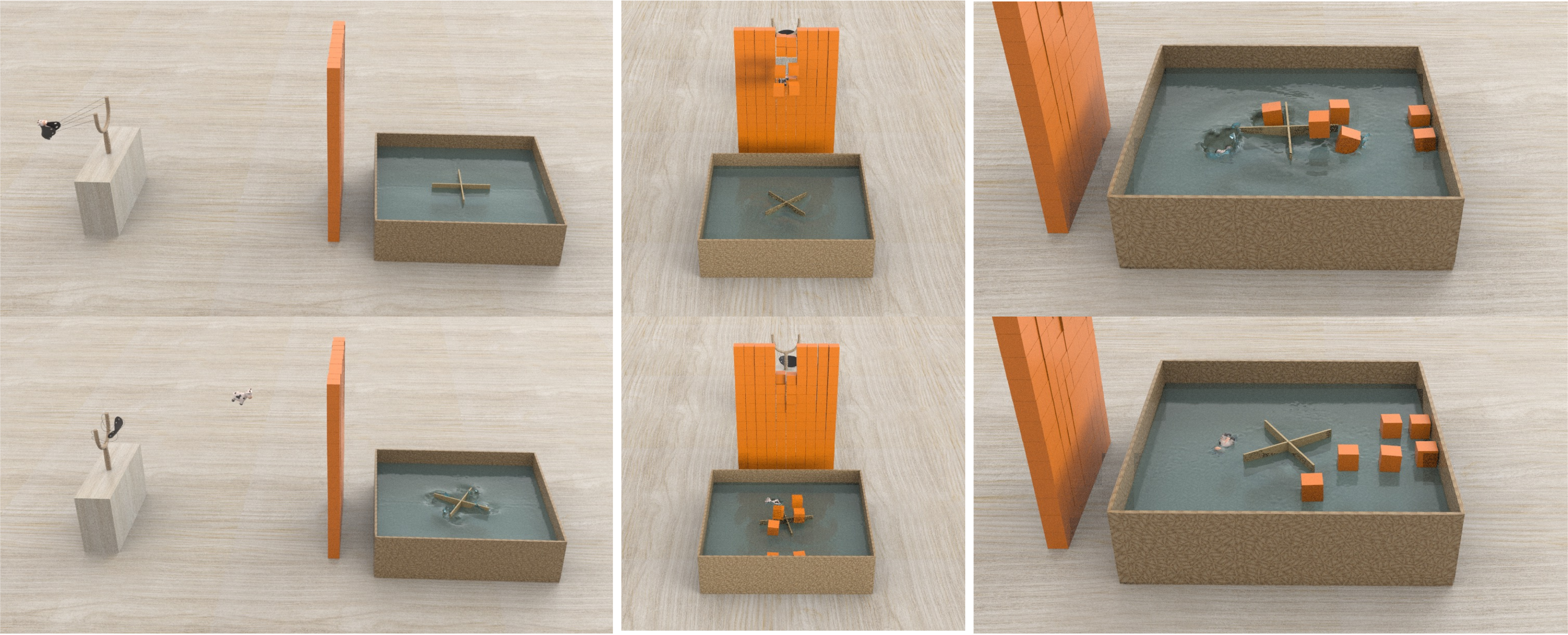}
  \caption{\textbf{Angry cow}. We show our method can simulate the coupling of materials in arbitrary codimensions, including fluid particles, rods (the rubber bands), thin shells (the leather pad), deformable solids (the cow), and rigid bodies (the cubes). We launch an angry cow with a slingshot, and the cow hits through the wall and then falls into the water. Interactions between various materials are all accurately captured.}
  \label{fig: angry cow}
\end{figure*}

\subsubsection{Fluid Dynamics}
While most existing SPH fluid solvers focus on incompressible fluids, our formulation treats fluids as weakly compressible, allowing us to couple fluids with deformable solids in a unified framework. We run a dam break simulation to compare our method with two SPH fluid solvers IISPH \cite{ihmsen2013implicit} and DFSPH \cite{bender2015divergence}. These methods typically use particle resampling \cite{akinci2012versatile, akinci2013coupling} or implicit representation \cite{koschier2017density, bender2019volume} to exert boundary counter-forces. Our method instead employs IPC \cite{li2020incremental} for more robust solid-fluid coupling, with penetration-free guarantee. We uniformly enforce the same CFL condition for all methods along with an upperbound at $5ms$, and use the volume map \cite{bender2019volume} for their boundary handling.  As shown in Fig. \ref{figure: fluid dynamics comparison}, though our formulation does not strictly enforce incompressibility, it produces natural fluid dynamics without visually observable volume loss. On the other hand, our method (0.45 min/frame) is slower than IISPH (0.31 min/frame) and DFSPH (0.15 min/frame) due to the more sophisticated boundary handling strategy. However, our proposed approach can couple SPH fluids and elastic solids with arbitrary constitutive models, while most existing SPH methods \cite{peer2018implicit, kugelstadt2021fast} treat elastic solids as incompressible, which is not generally applicable.

\subsubsection{Solid-Fluid Coupling}
We then compare our method with ElastoMonolith \cite{takahashi2022elastomonolith}, which couples Eulerian fluids with Lagrangian solids in a monolithic manner. Following their experiment setting, we run two solid-fluid coupling simulations with identical parameters using our method (Fig. \ref{fig: ElastoMonolith comparison}). The timing of our method for these two scenes are 24.1 sec/frame and 62.8 sec/frame respectively, both of which are over $5\times$ faster than ElastoMonolith according to their reported timings (253.2 sec/frame and 352.0 sec/frame). Coupling Eulerian fluids with Lagrangian solids requires dealing with geometric differences and it is often needed to perform SPD reformulation to make the linear system tractable. As stated in ElastoMonolith, this SPD reformulation can introduce many additional non-zeros to the system, especially when contacts are rich and solids are intricately shaped. Conversely, our method treats solids and fluids from a unified Lagrangian viewpoint, where solid-solid and solid-fluid contacts are resolved in a unified manner.

\subsection{Complex Scenarios}
We then evaluate the efficiency and robustness of our method in more complicated scenarios. We demonstrate our method correctly captures the buoyancy behavior in Fig.~\ref{fig: buoyancy}. Viscous fluids can also be naturally simulated (Fig.~\ref{fig: cream}), even with adjustable boundary friction (Fig.~\ref{fig: varying friction}). In addition, two-way coupling with thin shells (Fig.~\ref{fig: twist cylinder} and Fig.~\ref{fig: kick}) is also well supported with penetration-free guarantee. In Fig.~\ref{fig: angry cow}, we show that our framework can even simulate a tightly coupled system of geometries in arbitrary codimensions (0, 1, 2, and 3). Detailed parameter settings can be found in the appendix.

\section{Conclusion}
We presented a unified two-way strong coupling framework for weakly-compressible SPH fluids and nonlinear elastic FEM solids. To achieve this, we modeled solid-fluid interactions as contact forces between SPH particles and FEM boundary elements, applying IPC for guaranteed non-penetration and stability. As we track the volume change of SPH particles in an updated Lagrangian fashion, the incompressibility energy stays quadratic and nice particle distributions are maintained. Utilizing a symmetric approximation of discrete viscosity forces, we proposed a viscosity potential that fitted into optimization time integration. We then proposed a time splitting scheme with a contact proxy to efficiently solve the time integration optimization while maintaining robustness. The performance is further boosted by our matrix-free conjugate gradient method and a domain-decomposed solver based on Schur complement.

Compared to existing works \cite{zarifi2017positive, takahashi2022elastomonolith} coupling Eulerian fluids with Lagrangian elastic solids, our method treats both fluids and solids in a Langrangian manner, avoiding the need to handle different spatial discretizations. Under such a unified view, our method achieves more convenient and robust two-way coupling, even between fluids and codimensional solids. Likewise, different from existing SPH methods \cite{kugelstadt2021fast, peer2018implicit} that treat all materials as SPH particles, our formulation enjoys both the efficiency of SPH fluids and the accuracy of FEM solids.

There are many meaningful future research directions. 
First, when fluid DOFs dominate, building and querying the spatial hash for each fluid particle can become a considerable cost. In fact, since there is no solid-fluid contact for interior particles, we can construct the spatial data structure only in the intersection between the extended bounding boxes of the fluids and each solid for better efficiency. 
In addition, the adhesion between solids and fluids is also an interesting behavior to model. Similar to the barrier energy, adhesion forces can be exerted on close solid-fluid primitive pairs but in the opposite direction. 
Modeling adhesion via resolving the surface tension of fluids is also an interesting future work.
\bibliographystyle{ACM-Reference-Format}
\bibliography{reference}

\appendix


\newcommand{\norm}[1]{\left\lVert#1\right\rVert}






\section{Derivatives of Fluid Potentials}
The gradient and Hessian of the incompressibility potential w.r.t the fluid particle position are
\begin{equation}
\begin{aligned}
    \frac{\partial P_I(\mathbf{x})}{\partial \mathbf{x}} &= \sum_i k_I V_0 (J_i - 1) J^n_i h \frac{\partial (\nabla \cdot \mathbf{v}^{n+1}_i)}{\partial \mathbf{x}}, \quad\\
    \frac{\partial^2 P_I(\mathbf{x})}{\partial \mathbf{x}^2} &= \sum_i k_I V_0 (J^n_i)^2 h^2 \frac{\partial (\nabla \cdot \mathbf{v}^{n+1}_i)}{\partial \mathbf{x}} \left(\frac{\partial (\nabla \cdot \mathbf{v}^{n+1}_i)}{\partial \mathbf{x}}\right)^T.
\end{aligned}
\end{equation}
This constant Hessian matrix is obviously positive semi-definite (PSD) since it is simply the sum of the outer product of some vector with positive coefficients.

Similarly, the gradient and Hessian of the viscosity potential w.r.t the fluid particle position are
\begin{equation}
\begin{aligned}
    \frac{\partial P_V(\mathbf{x})}{\partial \mathbf{x}_i} &= \nu \sum_j \mathbf{V}_{ij} \mathbf{v}^{n+1}_{ij}, \quad
    \frac{\partial^2 P_V(\mathbf{x})}{\partial \mathbf{x}_i \partial\mathbf{x}_j} &= \left \{
            \begin{array}{ll}
                \frac{\nu}{\hat{h}} \sum_k \mathbf{V}_{ik}, j = i\\
                -\frac{\nu}{\hat{h}} \mathbf{V}_{ij}, j \neq i
            \end{array} \right..
\end{aligned}
\end{equation}
Since $ \mathbf{V}_{ij} = -4 (d+2) \frac{m_i m_j}{\rho_i + \rho_j}\frac{\nabla_i W_{ij} (\mathbf{x}^n_{ij})^T}{\|\mathbf{x}_{ij}^n\|^2 + 0.01\hbar ^2} \propto \mathbf{x}^n_{ij}(\mathbf{x}^n_{ij})^T$ is a $3\times3$ constant PSD matrix for any particle pair within a time step, the Hessian of the viscosity potential is a constant PSD matrix as well.

\section{Time Splitting}
\subsection{Baseline Time Splitting}
Applying time splitting, we can split the original time integration into a \textbf{fluid phase}
\begin{equation}
    \begin{cases}
        \ \mathbf{v}_f^{n+1/2} = \mathbf{v}_f^n + h \mathbf{M}_f^{-1} (-\nabla_f P([(\mathbf{x}_f^{n+1/2})^T,(\mathbf{x}_s^n)^T]^T) + \mathbf{f}_f)\\
        
        \ \mathbf{x}_f^{n+1/2} = \mathbf{x}_f^n + h \mathbf{v}_f^{n+1/2}
    \end{cases}
\end{equation}
and a \textbf{solid-coupling phase}
\begin{equation}
    \begin{cases}
        \ \mathbf{v}^{n+1} = 
        \begin{bmatrix}
            \mathbf{v}_f^{n+1/2} \\
            \mathbf{v}_s^{n}
        \end{bmatrix} + h \mathbf{M}^{-1} (-\nabla \Psi(\mathbf{x}^{n+1}) -\nabla C(\mathbf{x}^{n+1}) + 
        \begin{bmatrix}
            \mathbf{0} \\
            \mathbf{f}_s
        \end{bmatrix})\\
        
        \ \mathbf{x}^{n+1} = \mathbf{x}^n + h \mathbf{v}^{n+1}
    \end{cases},
\end{equation}
where $\mathbf{f}_f$ and $\mathbf{f}_s$ are the external forces on the fluids and the solids respectively. The two phases of this baseline time splitting scheme have equivalent optimization forms
\begin{equation}
\boxed{
        \begin{aligned}
            \mathbf{x}^{n+1/2}_f &= arg\min_{\mathbf{x}_f} \frac{1}{2}\| \mathbf{x}_f - \hat{\mathbf{x}}^n_f\|^2_{\mathbf{M}_f} + h^2 P([\mathbf{x}_f^T,(\mathbf{x}_s^n)^T]^T)\\
            \mathbf{x}^{n+1} &= arg\min_\mathbf{x} \frac{1}{2}\|\mathbf{x} - \hat{\mathbf{x}}^{n+1/2}\|^2_\mathbf{M} + h^2 (\Psi(\mathbf{x}) + C(\mathbf{x}))
        \end{aligned}
}
\end{equation}
where $\hat{\mathbf{x}}^n_f = \mathbf{x}^n_f + h \mathbf{v}^n_f + h^2 \mathbf{M}_f^{-1} \mathbf{f}_f$ and $\hat{\mathbf{x}}^{n+1/2} = \mathbf{x}^n + h [(\mathbf{v}^{n+1/2}_f)^T, (\mathbf{v}^n_s)^T]^T + h^2 \mathbf{M}^{-1} [\mathbf{0}^T, \mathbf{f}_s^T]^T$ with $\mathbf{v}^{n+1/2}_f = (\mathbf{x}^{n+1/2}_f - \mathbf{x}_f^n)/h$. The details of the optimization algorithm can be found in Alg.~\ref{alg: simple time splitting}.

\begin{algorithm}
\caption{Baseline Time Splitting}\label{alg: simple time splitting}
\begin{algorithmic}[1]
\State $\mathbf{x} \gets \mathbf{x}^n, \hat{\mathbf{x}}^n_f \gets \mathbf{x}^n_f + h \mathbf{v}^n_f + h^2 \mathbf{M}^{-1}_f \mathbf{f}_f$
\State $\text{SPH Neighbor Search } \& \text{ Density Update}$
\State // Fluid Phase
\State $\mathbf{H}_f \gets h^2\nabla^2_f P([\mathbf{x}_f^T, (\mathbf{x}_s^n)^T]) + \mathbf{M}_f$
\State $\mathbf{p}_f \gets -\mathbf{H}^{-1}_f\left(h^2\nabla_f P([\mathbf{x}_f^T, (\mathbf{x}_s^n)^T]^T) + \mathbf{M}_f(\mathbf{x}_f - \hat{\mathbf{x}}^n_f)\right) $
\State $\mathbf{x}_f \gets \mathbf{x}_f + \mathbf{p}_f$
\State $\mathbf{x}^{n+1/2}_f \gets \mathbf{x}_f, \mathbf{v}^{n+1/2}_f = (\mathbf{x}_f - \mathbf{x}_f^n) / h$
\State // Solid Coupling Phase
\State $\hat{\mathbf{x}}^{n+1/2} \gets \mathbf{x}^n + h [(\mathbf{v}^{n+1/2}_f)^T, (\mathbf{v}^n_s)^T]^T + h^2 \mathbf{M}^{-1} [\mathbf{0}^T, \mathbf{f}_s^T]^T$
\Do
    \State $\mathbf{H} \gets h^2 \left(\nabla^2 \Psi(\mathbf{x}) + \nabla^2 C(\mathbf{x})\right) + \mathbf{M}$
    \State $\mathbf{p} \gets -\mathbf{H}^{-1}\left(h^2(\nabla \Psi(\mathbf{x}) + \nabla C(\mathbf{x})) + \mathbf{M}(\mathbf{x} - \hat{\mathbf{x}}^{n+1/2})\right) $
    \State $\alpha \gets \text{Backtracking Line Search with CCD}$
    \State $\mathbf{x} \gets \mathbf{x} + \alpha \mathbf{p}$
\doWhile{$\frac{1}{h} \|\mathbf{p}\| > \epsilon$}
\State $\mathbf{x}^{n+1} \gets \mathbf{x}, \mathbf{v}^{n+1} \gets (\mathbf{x} - \mathbf{x}^n)/h$\\
\Return $\mathbf{x}^{n+1}, \mathbf{v}^{n+1}$
\end{algorithmic}
\end{algorithm}

\subsection{Error Analysis}
The position update of implicit Euler and our proxy-enhanced time splitting scheme can be respectively expressed as
\begin{equation}
\begin{aligned}
    \mathbf{x}^{n+1}_\text{IE} & = \mathbf{x}^n + h\mathbf{v}^n + h^2 \mathbf{M}^{-1} \Bigl(-\nabla P(\mathbf{x}^{n+1}_\text{IE}) \\
        &\qquad- \nabla \Psi(\mathbf{x}^{n+1}_\text{IE}) - \nabla \csf (\mathbf{x}^{n+1}_\text{IE}) - \nabla \css (\mathbf{x}^{n+1}_\text{IE}) + \mathbf{f}_\text{ext} \Bigr)\\
    \mathbf{x}^{n+1} & = \mathbf{x}^n + h\mathbf{v}^n + h^2 \mathbf{M}^{-1} \Bigl(-\nabla P(\mathbf{x}^{n+1/2}) - \nabla \Psi(\mathbf{x}^{n+1}) \\
        &\qquad- \nabla \hcsf (\mathbf{x}^{n+1/2}) -\frac{1}{2}\nabla \csf(\mathbf{x}^{n+1}) - \nabla \css(\mathbf{x}^{n+1}) + \mathbf{f}_\text{ext}\Bigr).
\end{aligned}
\end{equation}
If we define
\begin{equation}
\begin{aligned}
    e(\mathbf{x}) &= \Big\lVert \mathbf{x}^n + h\mathbf{v}^n + h^2 \mathbf{M}^{-1} \Bigl(-\nabla P(\mathbf{x}) - \nabla \Psi(\mathbf{x}) \\
        & \qquad - \nabla \csf (\mathbf{x})  - \nabla \css (\mathbf{x}) + \mathbf{f}_\text{ext}\Bigr) - \mathbf{x} \Big\rVert,
\end{aligned}
\end{equation}
$\mathbf{x}^{n+1}_\text{IE}$ given by implicit Euler satisfies $e(\mathbf{x}^{n+1}_\text{IE})= 0$, while for $\mathbf{x}^{n+1}$ from our scheme, we have $e(\mathbf{x}^{n+1}) = \mathcal{O}(h^4)$.
Specifically,
\begin{equation}
\begin{aligned}
    e(\mathbf{x}^{n+1}) &= h^2 \Big\lVert \mathbf{M}^{-1} (\nabla P(\mathbf{x}^{n+1/2}) - \nabla P(\mathbf{x}^{n+1}) \\
        &\qquad \qquad+\nabla \hcsf (\mathbf{x}^{n+1/2}) - \frac{1}{2}\nabla \csf(\mathbf{x}^{n+1})) \Big\rVert \\
        &= \mathcal{O}\Bigg( h^3 \Big\lVert \mathbf{M}^{-1} \Bigl(\nabla^2 P(\mathbf{x}^n)(\mathbf{v}^{n+1/2} - \mathbf{v}^{n+1}) + \\
        &\qquad \qquad \frac{1}{2}\nabla^2 \csf(\mathbf{x}^n)(\mathbf{v}^{n+1/2} - \mathbf{v}^{n+1})\Bigr) \Big\lVert \Bigg)\\
        &= \mathcal{O}\Bigg( h^4 \Big\lVert \mathbf{M}^{-1} \Bigl(\nabla^2 P(\mathbf{x}^n) + \frac{1}{2} \nabla^2 \csf(\mathbf{x}^n) \Bigr) \mathbf{M}^{-1} \Big(\nabla \Psi(\mathbf{x}^n) \\
        &\qquad \qquad + \frac{1}{2} \nabla C_\text{sf}(\mathbf{x}^n) + \nabla C_\text{ss}(\mathbf{x}^n) \Big) \Big\rVert \Bigg) \\
        &= \mathcal{O}(h^4).
\end{aligned}
\end{equation}
Here we assume that, in our discretized domain, the distance between any pair of primitives (particle-particle pair, particle-triangle pair and triangle-triangle pair) has a lower bound $\epsilon$. Thus $\nabla^2 P(\mbf{x}^n)$, $\nabla^2 C_\text{sf}(\mbf{x}^n)$, $\nabla \Psi(\mbf{x}^n)$, $\nabla C_\text{sf}(\mbf{x}^n)$ and $ \nabla C_\text{ss}(\mbf{x}^n)$ are all bounded, and this indicates our method has an $\mathcal{O}(h^4)$ difference compared to implicit Euler solution. 
Since implicit Euler has an $\mathcal{O}(h^2)$ error compared to the PDE solution, our proposed time splitting scheme shares the same order of accuracy with implicit Euler when it is stable.

\section{Experiment of Complex Scenarios}
In this section, we describe the experiment settings of our simulations in various complex scenarios and briefly discuss the results.
\paragraph{Buoyancy}
We drop three elastic elephants with varying densities into the water (1000 $kg/m^3$). The light grey elephant (200 $kg/m^3$) floats on the surface; the blue elephant (700 $kg/m^3$) is around half immersed in the water; and the red elephant (1200 $kg/m^3$) sinks into the bottom. This demonstrates that our method correctly captures the buoyancy behavior.

\paragraph{Varying Friction}
We drop three viscous bunnies onto the slope with different coefficients of friction (orange bunny: 0.5, green bunny: 0.03, blue bunny: 0). All three bunnies share the same dynamic viscosity coefficients 100 $kg/m^3$ and the angle of slope is $30^{\circ}$.

\paragraph{Twist Cylinder}
Coupling fluids with thin shells is challenging since penetration can easily happen without careful treatments. As stated in \cite{zarifi2017positive}, Eulerian fluids may flow through solids if their thickness is less than a grid cell size. Conversely, our approach adopts a unified Lagrangian view and penetration-free is guaranteed by IPC. In this example, we simulate twisting a cylinder full of water. The cylinder is modeled as a thin shell with a $2mm$ thickness, and there are two holes in the front and back sides of this cylinder respectively. The left side and right side are rotated at $72^{\circ}/s$ and are slowly moved towards each other at $2 cm/s$. As we twist the cylinder, the water gets squeezed out through the holes. This simulation demonstrates our method produces stable simulation results with penetration-free guarantee.

\paragraph{Cream}
This example exhibits the coupling behaviors of viscous fluids and elastic solids. We use an elastic spoon to stir the cream in a porcelain bowl. The spoon handle rotates around y-axis at $360^{\circ}/s$ ($0.2m/s$) while the bowl is fixed at the table. As shown in our simulation results, the spoon gets deformed due to the resistance forces it receives from the viscous cream while stirring. 

\paragraph{Angry Cow} We then show our framework can simulate natural physical behaviors of geometries in arbitrary codimensions (0, 1, 2, and 3) as well as their interactions. In this scene, the codimensional-0,1,2 objects respectively refer to fluid particles, rubber bands and the leather pad. A deformable cow is launched by the slingshot, hitting the wall consisting of rigid cubes, and then falling into the water pool, producing interesting physical behaviors. The density of the rigid cubes and the cow are $100 kg/m^3$ and $700 kg/m^3$ respectively.

\paragraph{Kick Water} In this example, we show a scene where a mannequin dressed in a multilayer skirt kicks in a large water pool, involving complex interactions between fluid particles and garments. As the mannequin moves in the water, our method produces natural deformation of the skirt caused by the contact with water; as it kicks out of the water at a high speed, the resulting water splash is also correctly captured. Our method well resolves the contacts among fluids particles, thin garments and rapidly moving complex boundaries with penetration-free guarantee.



\end{document}